\documentclass[twocolumn,showpacs,preprintnumbers,amsmath,amssymb,floatfix]{revtex4}

\usepackage{multirow,amssymb,amsbsy,amsmath}
\usepackage{graphicx}
\usepackage{dcolumn}
\usepackage{bm}
\usepackage{color}
\usepackage{ulem}

\begin{document}
\title{Modeling and control of entanglement dynamics in laser cooling of trapped atoms}
\author{Maryam Roghani}
\email{maryam.roghani@physik.uni-hamburg.de}
\author{Heinz-Peter Breuer}
\email{breuer@physik.uni-freiburg.de}
\author{Hanspeter Helm}
\email{helm@uni-freiburg.de}

\affiliation{Physikalisches Institut, Universit\"at Freiburg,
Hermann-Herder-Strasse 3, D-79104 Freiburg, Germany}

\date{\today}

\begin{abstract}
We discuss the dynamical behavior of the entanglement between the internal and the external degrees of freedom of a trapped atom in electromagnetically-induced transparency (EIT) laser cooling. It is shown that essential features of the intricate entanglement dynamics observed in full numerical simulations of the underlying quantum master equation can be understood in terms of a two-state model on the basis of Landau-Zener splittings in the atom-laser field Hamiltonian. An extension of this model to an effective non-Hermitian Hamiltonian is constructed which describes the decay of entanglement by spontaneous emission processes. We also discuss schemes for the control of entanglement and demonstrate that a permanent entanglement can be imprinted on trapped atoms through a rapid switch off of the driving fields. Finally, we point out fundamental distinctions between the entanglement created in EIT cooling and in the cooling scheme based on velocity-selective coherent population trapping.
\end{abstract}


\maketitle

\section{Introduction}
Entanglement is a characteristic property of the states of composite quantum systems and represents a key feature for many proposed applications of quantum mechanics. Since quantum systems are often strongly affected by decoherence through the interaction with an environment \cite{BP}, the dynamics of entanglement in the presence of dissipative processes plays an important role in many contexts \cite{ABU}. One of the challenges is, for example, to find ways to protect the coherence of quantum states against external noise and to maintain entanglement for an extended period of time.

Building on the ideas presented in \cite{MRPRL} we investigate here in detail the dynamics of the entanglement that arises between the internal electronic and the external translational degrees of freedom of trapped atoms in laser cooling. The cooling scheme investigated employs the cooling of the vibrational motion of the atom in an external potential by means of electromagnetically-induced transparency (EIT) using two counter-propagating and suitable tailored laser fields. In EIT cooling entanglement is a fundamental requisite for cooling, but fades once cooling into the vibrational ground state is achieved. The importance of entanglement for the cooling process permits to imprint permanent entanglement between the atom's internal and external degrees of freedom.

The situation studied here is reminiscent of the cooling mechanism based on velocity-selective coherent population trapping (VSCPT)
which has been considered by Arimondo \cite{Arimondo00} who studied the entanglement between momentum eigenstates of the free atom and its internal electronic degrees of freedom. By contrast, here we study the entanglement between vibrational eigenstates in the external potential and the atom's internal electronic degrees of freedom. In the ultimate cooling limit, the degree of entanglement reached is very different from that in VSCPT. In the latter case a maximally entangled state appears when the translational motion is fully cooled \cite{Arimondo00}.

The degree of entanglement in the atomic density matrix $\rho(t)$ will be quantified here by means of the negativity \cite{VIDAL} which is defined as
\begin{equation} \label{mr-5}
 {\mathcal{N}}(t) = {\mathcal{N}}(\rho(t)) = \frac{1}{2}\Big(||\rho^{\rm T}(t)||-1\Big).
\end{equation}
Here, $\rho^{\rm T}(t)$ represents the partial transpose of
$\rho(t)$ with respect to either the internal or the external degrees of freedom, and $||A||={\mathrm{Tr}}|A|$ denotes the trace norm of an operator $A$. The negativity ${\mathcal{N}}(\rho)$ is a nonnegative, convex function and represents a useful measure for the amount of entanglement of quantum states which is easy to compute. It represents an entanglement monotone, vanishes for separable states and takes on a maximum value if $\rho$ is a pure, maximally entangled state. Performing extensive numerical simulations of the underlying quantum master equation that describes the dynamics of the atomic density matrix, we are able to study the complex time evolution of the negativity in EIT cooling.

The paper is organized as follows. After a brief introduction of the EIT-cooling scheme and of the quantum master equation used to model this scheme, we present in Sec.~\ref{theory} the intricate dynamical behavior of the negativity which is observed during the vibrational cooling process \cite{MRPRL} and show that the basic features of the entanglement dynamics can be traced to Landau-Zener splittings (avoided crossings) \cite{LANDAU,ZENER,STUECKELBERG} in the spectrum of the atom-laser field Hamiltonian. We further construct an effective Hamiltonian that describes the behavior of entanglement under dissipation induced by spontaneous emission processes. In Sec.~\ref{numerics} we present results of numerical simulations of the full master equation of EIT cooling, and compare these with the predictions derived from the Landau-Zener model. Furthermore, we discuss here the control of entanglement through sudden and adiabatic switching off the driving laser fields. Finally, we draw our conclusions in Sec.~\ref{conclu} and discuss in some detail the differences in the entanglement dynamics of the cooling schemes of EIT and VSCPT.

\section{Theory}\label{theory}

\subsection{Theoretical model and master equation}
In a trapped atom system the Lamb-Dicke parameter
$\eta=k/\sqrt{{2m\omega}}$ ($\hbar$=1) controls the probability of changing the vibrational quantum number $n$ in electronic
transitions \cite{You}. For small $\eta$ the photon-recoil energy
$k^2/2m$ is an inconsiderable fraction of the trap vibrational
frequency $\omega$. In the Lamb-Dicke regime, $\eta\sqrt{n+1}\!\ll\! 1$, electronic transitions with a change of the vibrational level are active, primarily in the first sideband transitions, $\Delta n=\pm 1$, their probability being proportional to $(n\!+\!1)\eta^2$, where $n$ is the vibrational quantum number.
A rapid removal of vibrational energy of a trapped atom occurs for suitable experimental parameters \cite{Morigi,roos, RBH}, i.e., a difference of ac-Stark shifts of the two ground states of the order of $\omega$. Under these conditions, with blue detuning of two counter-propagating lasers, a vibrationally excited atom can be driven into states with mean values $\langle n \rangle$ near zero, as has been demonstrated in experiments with atomic ions \cite{schmidt-kaler}.

\begin{figure}[h]
\includegraphics[width=0.55\columnwidth]{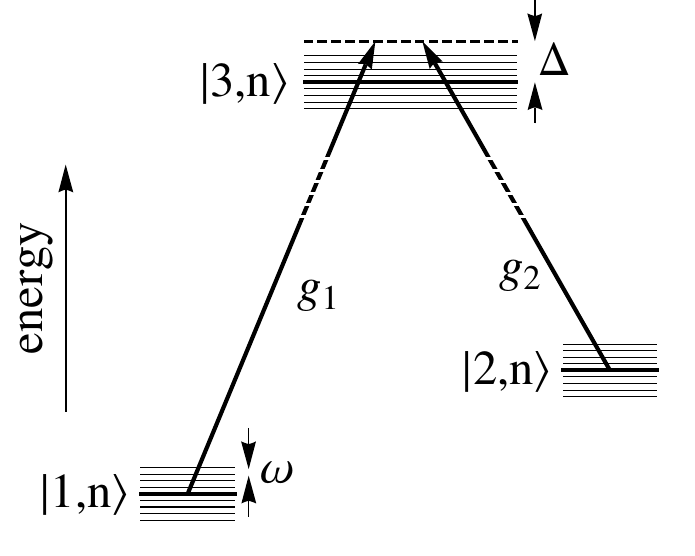}
\caption{Level scheme of the trapped atom.}
\label{MR-Figure-1}
\end{figure}

We have explored this concept in a model based on the full quantum master equation
\begin{equation} \label{mr-1}
 \frac{d}{dt}\rho(t) = -{\rm{i}}[H^{\eta},\rho(t)] + {\mathcal{L}}\rho(t)
\end{equation}
for the interaction picture density matrix $\rho(t)$ representing
the combined state of vibrational and electronic degrees of
freedom of the trapped atom. The total Hamiltonian
\begin{equation} \label{mr-2}
 H^{\eta} = H_{\rm cm} + H_{\rm el} + H^{\eta}_{\rm int}
\end{equation}
of the model consists of three parts: $H_{\rm cm}=\omega
a^{\dagger}a$ describes the vibrational degree of freedom of a
harmonically trapped atom with raising and lowering operators
$a^{\dagger}$ and $a$, and $H_{\rm el}=\Delta (|1\rangle\langle 1|
+ |2\rangle\langle 2|)$ represents the electronic degree of
freedom with detuning $\Delta$. The electronic states are denoted
by $|i\rangle$, $i=1,2,3$, and form a $\Lambda$-type level
structure sketched in Fig.~\ref{MR-Figure-1}.
The Hamiltonian
\begin{equation} \label{mr-3}
 H^{\eta}_{\rm int} = \frac{g_1}{2} e^{ikx}|3\rangle\langle 1|
 + \frac{g_2}{2} e^{-ikx}|3\rangle\langle 2| + \rm{h.c.}
\end{equation}
models the interaction with two counter-propagating laser beams,
the most efficient configuration for cooling, where $x$ denotes
the atomic position and $k$ the wavenumber ($k_1 \approx k_2 =
k$). Note that $H^{\eta}_{\rm int}$ depends on the Lamb-Dicke
parameter $\eta$ through $kx=\eta(a^{\dagger}+a)$.

For EIT cooling the detunings are chosen positive
\begin{equation}
\Delta_1=\omega_{L1}-\omega_{31} >0, \qquad
\Delta_2=\omega_{L2}-\omega_{32}>0,
\label{mr-6}
\end{equation}
and equal, $\Delta_1=\Delta_2=\Delta$. The latter condition implies that in the dressed state frame the non-interacting states $|1\rangle$ and $|2\rangle$ are at equal total energy and separated from the excited state $|3\rangle$ by the detuning $\Delta$.

Finally, dissipation caused by spontaneous emissions is modeled by the Lindblad superoperator (see, e.g., Ref.~\cite{BP})
\begin{equation}
\label{mr-4}
 {\mathcal{L}}\rho = \sum_{j=1,2} \sum_{q=\pm} \frac{\Gamma}{4}
 \left[ \sigma_{jq}\rho \sigma^{\dagger}_{jq} - \frac{1}{2}
 \left\{ \sigma_{jq}^{\dagger}\sigma_{jq}, \rho \right\} \right],
\end{equation}
where $\Gamma$ denotes the spontaneous emission rate and
$\sigma_{j+} = |j\rangle\langle 3| e^{ikx}$, $\sigma_{j-} =
|j\rangle\langle 3| e^{-ikx}$. The Lindblad operators
$\sigma_{j+}$ describe the transitions from the excited electronic
state $|3\rangle$ to the ground states $|j=1,2\rangle$ with a
photon emitted parallel to the harmonic oscillator axis, while the
$\sigma_{j-}$ describe the transitions from $|3\rangle$ to
$|j=1,2\rangle$ with a photon emitted counter-propagating to the
harmonic oscillator axis. This is a good compromise
short of doing a full 3D simulation as the recoil of the
translational motion is zero for emissions perpendicular to the
trap axis, while it leads to Doppler heating or cooling for
emissions along the trap axis.

\begin{figure}[h]
\includegraphics[width=0.73\columnwidth]{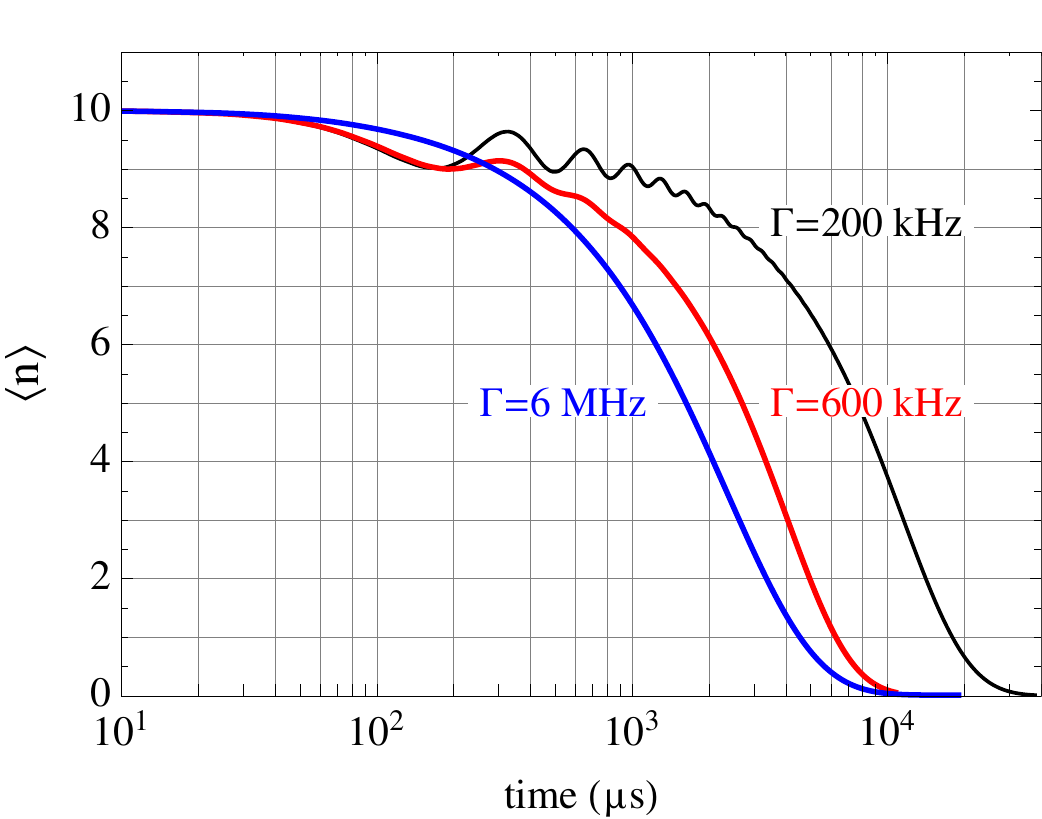}\\
\vspace{2mm}
\includegraphics[width=0.77\columnwidth]{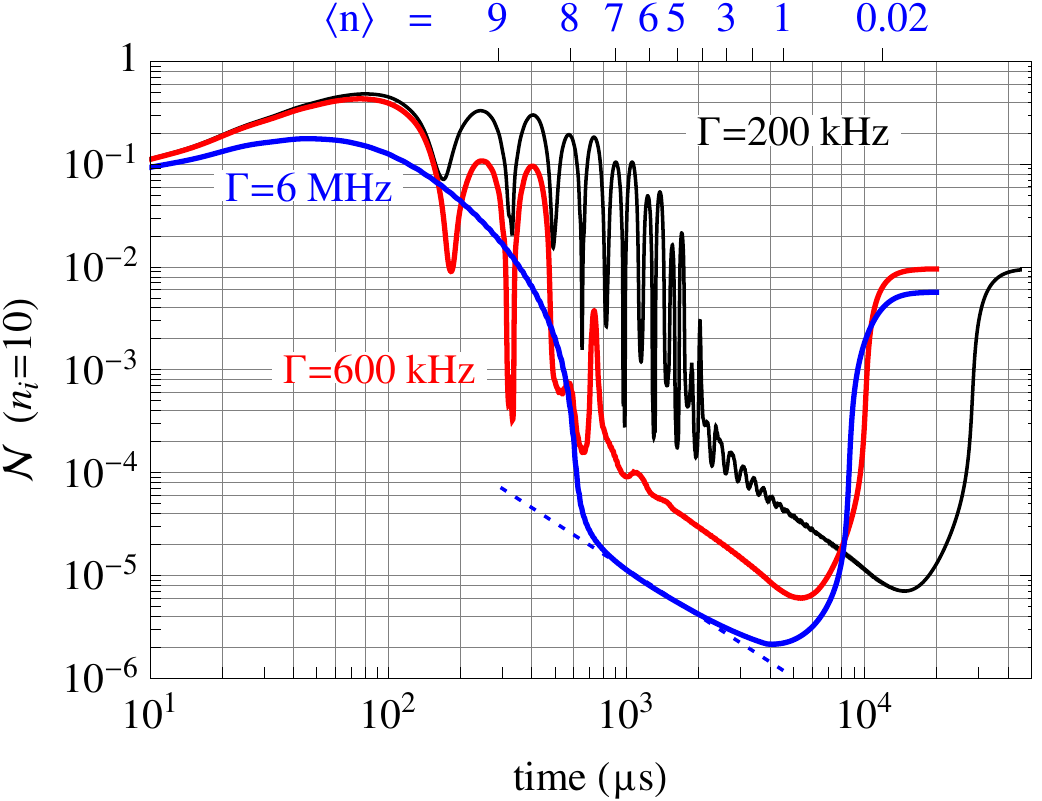}
\caption{\label{MR-Figure-2} a) Dynamics of vibrational cooling for the initial vibrational level $n_i=10$ for $\eta=0.1$ and for three values of the excited state width $\Gamma$. b) Dynamics of the negativity during the cooling process. The top scale gives
the mean vibrational quantum number for the 6 MHz result. The
dotted line represents a power law decay, ${\mathcal{N}}\propto
t^{-3/2}$. The parameters are $g_1\!=\!1.34$, $g_1/g_2\!=\!10$, $\omega\!=\!0.03$ and $\Delta\!=\!15$ in units of $2\pi$~MHz.}
\end{figure}

Typical results obtained by numerically solving Eq.~\eqref{mr-1} \cite{MRPRL,RBH} are given in Fig.~\ref{MR-Figure-2}a which gives the dependence of the mean vibrational quantum number as a function of time for $\eta=0.1$, for three values of the excited state width $\Gamma$.
The analysis of the  radiation field which is scattered from the atom during the process of cooling \cite{RBH} showed that it is the interplay of stimulated Raman transitions between vibronic states and spontaneous emissions which controls the speed and degree of cooling.

As the Raman transitions are active in a space of inseparable wavefunctions it appeared useful to us to  also explore the temporal development of entanglement during the cooling process.
As apparent from  Fig.~\ref{MR-Figure-2}b an intricate behavior of the negativity is observed during such a vibrational cooling process. We see a rapid build-up of entanglement, a subsequent exponentially damped oscillation, followed by a precipitous drop and a power law behavior and, finally, a rebirth of entanglement towards the non-equilibrium stationary state.

We show below that this complex behavior is related to Landau-Zener splittings or avoided crossings in the spectrum of the atom-laser field Hamiltonian. The dominant source of entanglement is a near degeneracy of vibronic states, which appears as a consequence of the ac-Stark shift, suitably chosen for EIT-cooling. Two-photon transitions associated with a change in vibrational quantum number occur and lead to a periodic oscillation of entanglement, damped by spontaneous emission.

\subsection{Landau-Zener model for entanglement dynamics}\label{LZ-Model}

\subsubsection{Near degeneracies of vibronic states}

\begin{figure}[h]
\includegraphics [width=0.46\columnwidth]{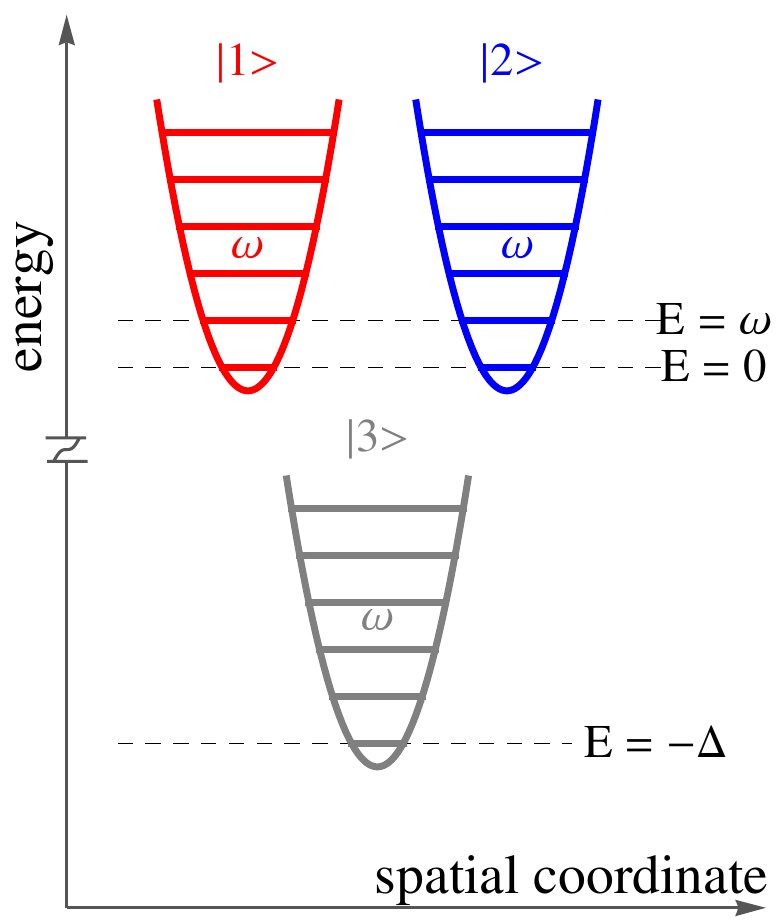}
\hspace{3mm}
\includegraphics [width=0.46\columnwidth]{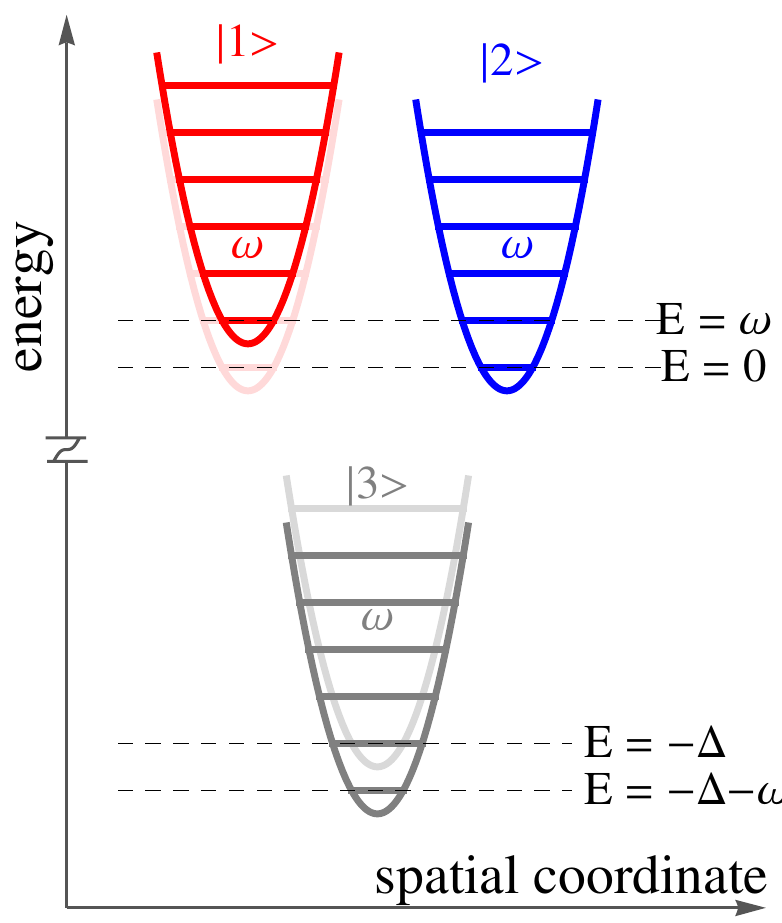}
\caption{\label{MR-Figure-3} The dressed atomic states in the absence of inter\-action for $g_1\!\gg\! g_2$ and an ac-Stark shift of magnitude $\omega$, displaced along the horizontal axis for clarity. }
\end{figure}

To introduce the Landau-Zener model we first neglect spontaneous emission and set the Lamb-Dicke parameter to zero. Then the interaction Hamiltonian is
\begin{equation} \label{mr-7}
 H^{0}_{\rm int} = \frac{g_1}{2} |3\rangle\langle 1|
 + \frac{g_2}{2} |3\rangle\langle 2| + \rm{h.c.}
\end{equation}
EIT-cooling is most efficient with regard to the cooling limit if one of the lasers is very much stronger than the other one, such that the difference of ac-Stark shifts experienced by the ground states is approximately equal to~$\omega$. This situation is illustrated in Fig.~\ref{MR-Figure-3}, which gives  the
dressed states of the harmonic oscillator, spatially separated for clarity. The assumption is made that laser one is very much stronger than laser two, thus the ac-Stark shift between $|1\rangle$ and $|3\rangle$ repels these two states while the shift of state $|2\rangle$ is negligible. For trapped neutral atoms we typically have $\Delta \gg \omega$, the separation between the dressed ground states and the excited state being much larger than it appears from Fig.~\ref{MR-Figure-3}.

In the dressed state picture, the eigenstates of the Hamiltonian $H_{\rm{el}}+H^{0}_{\rm{int}}$ which are given by
\begin{equation} \label{mr-8}
 \left(H_{\rm{el}}+H^{0}_{\rm{int}}\right)|\phi_i\rangle=\varepsilon_i |\phi_i\rangle
\end{equation}
have the energy eigenvalues
\begin{eqnarray}
\label{mr-9}
\varepsilon_1&\!=\!&+\frac{1}{2}\Big{(}\Omega\!+\!\Delta \Big{)} \\
\label{mr-10}
\varepsilon_2&\!=\!&+\Delta \\
\label{mr-11}
\varepsilon_3&\!=\!&-\frac{1}{2}\Big{(} \Omega-\Delta \Big{)}\, ,
\end{eqnarray}
where we have abbreviated $\Omega=\sqrt{g_1^2+g_2^2+\Delta^2}$.
The corresponding eigenvectors $|\phi_i\rangle$ are given by
\begin{eqnarray}
\label{mr-12}
|\phi_1\rangle&\!=\!&
\frac{1}{\sqrt{2 \Omega \, (\Omega-\Delta)}}\, \big(g_1 |1\rangle +g_2|2\rangle+(\Omega -\Delta)|3\rangle\big),  \\
\label{mr-13}
|\phi_2\rangle&\!=\!&\frac{1}{\sqrt{g_1^2+ g_2^2}} \, \big{(}-g_2 \, |1\rangle +g_1\, |2\rangle  \,  \big{)},  \\
\label{mr-14}
|\phi_3\rangle&\!=\!&
\frac{g_1}{\sqrt{2 \Omega \, (\Omega+\Delta)}}\, \big(g_1 |1\rangle +g_2|2\rangle-(\Omega +\Delta)|3\rangle\big).
\end{eqnarray}
With the choice $g_1\! \gg\! g_2$ we have $|\phi_2\rangle \!\approx \!|2\rangle$, while the state $|\phi_1\rangle$ carries a small contribution of the excited electronic state $|3\rangle$.

For vanishing Lamb-Dicke parameter, no coupling between electronic states exists via sideband transitions. In this case the three electronic states are always tied to a specific vibrational level. The eigenstates of the total system described by the Hamiltonian
\begin{equation}
 H^0=H_{\rm{cm}}+H_{\rm{el}}+H^{0}_{\rm{int}}
\end{equation}
are given by tensor products of the dressed electronic states $|\phi_i\rangle$ and the vibrational eigenstates $|n\rangle$,
\begin{equation}
 |\phi_i,n\rangle = |\phi_i\rangle \otimes |n\rangle,
\end{equation}
with the corresponding eigenvalues $\varepsilon_{i,n}=\varepsilon_i+n\omega$. When the ac-Stark shift is adjusted such that the two-photon detuning between Stark-shifted states equals the trap frequency,
\begin{equation} \label{mr-15}
 \omega = \frac{1}{2}(\Omega-\Delta),
\end{equation}
then the states $|\phi_1,n\!-\!1\rangle$ with energy $\varepsilon_{1,n-1}$ and $|\phi_2,n\rangle$ with energy $\varepsilon_{2,n}$ become degenerate, as is illustrated in Fig.~\ref{MR-Figure-3}.

\begin{figure}[t]
\includegraphics [width=0.48\columnwidth]{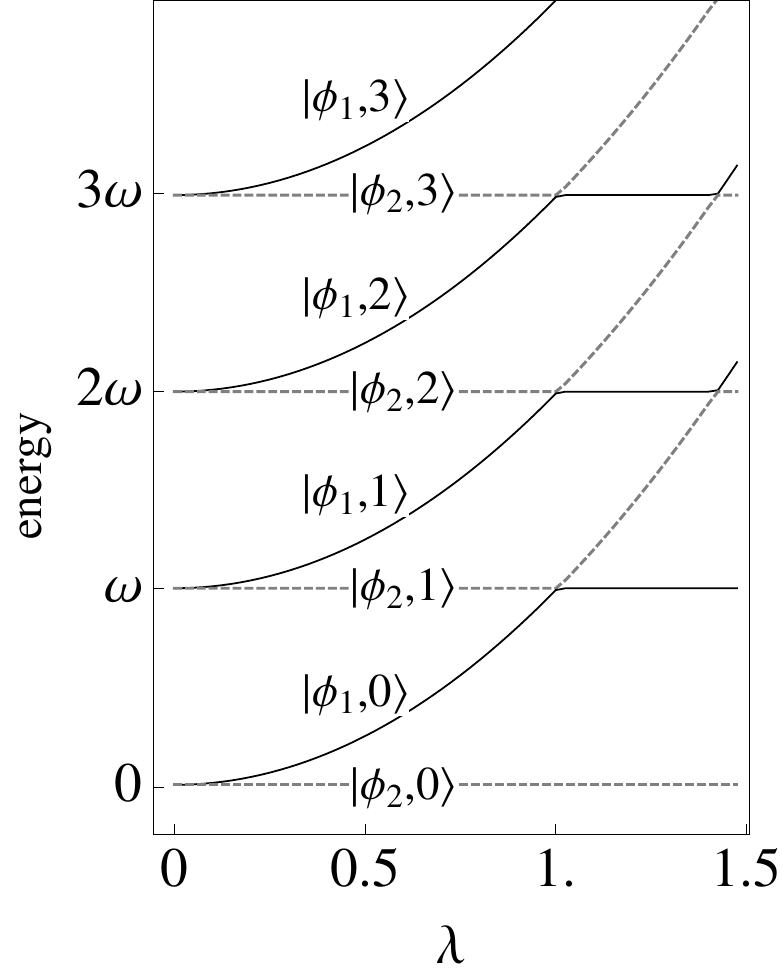}
\hspace{1mm}
\includegraphics [width=0.48\columnwidth]{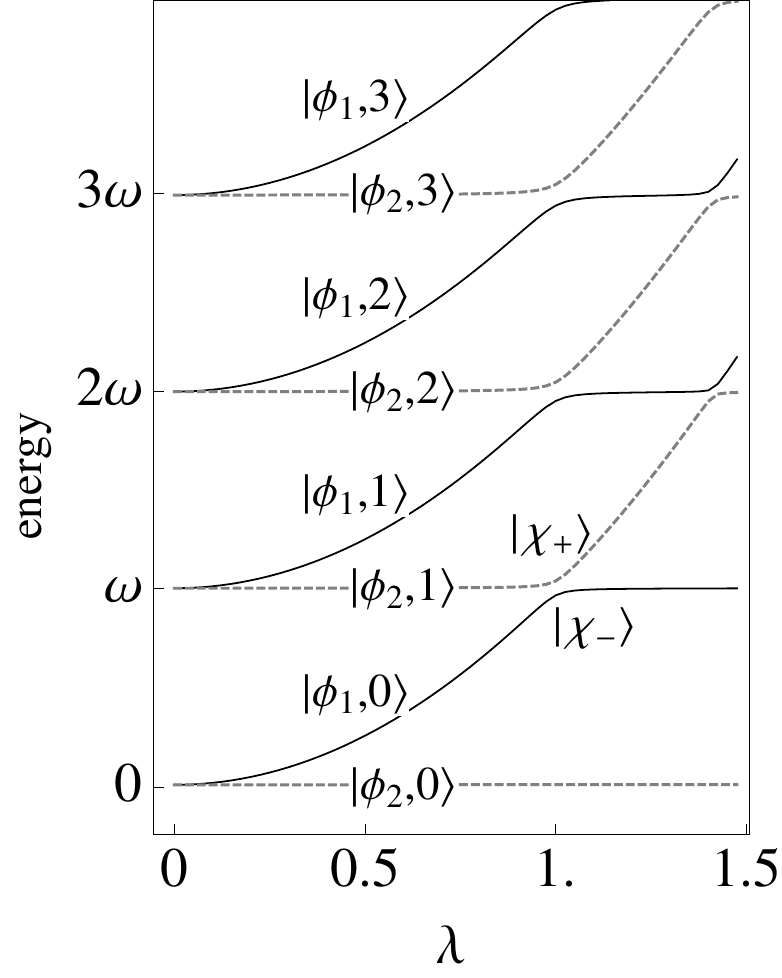}
\caption{\label{MR-Figure-4} Energy eigenvalues as a function of the parameter~$\lambda$.  The lowest Landau-Zener crossing appears for $\eta=0$ at $\lambda=1$. Energies are given as $\varepsilon_{i,n}-\Delta$.
Left: $\eta=0$,  right: interacting states for a Lamb-Dicke parameter $\eta=0.1 $ .}
\end{figure}

For a more general description we consider Rabi frequencies $g_i(\lambda)$ which vary linearly with the parameter $\lambda$, $g_1(\lambda)=\lambda \,g_1$ and $g_2(\lambda)=\lambda\, g_2$. Under the resonance condition (\ref{mr-15}) the energies of the pairs of states $|\phi_1,n\!-\!1\rangle$ and $|\phi_2,n\rangle$ then cross at $\lambda=1$, as is shown in the energy diagram of Fig.~\ref{MR-Figure-4}. It is known from the Landau-Zener scenario that this exact degeneracy is lifted under a generic perturbation. In fact, for a nonvanishing Lamb-Dicke parameter, $\eta>0$, the levels exhibit an avoided crossing (Landau-Zener splitting) with a certain energy gap $\Delta E$, as is shown in Fig.~\ref{MR-Figure-4}b.

To estimate the energy gap $\Delta E$ we consider the first order approximation of the interaction Hamiltonian $H^{\eta}_{\rm int}$ which is given by
\begin{equation} \label{mr-16}
 H^{(1)}_{\rm{int}}=\frac{\rm{i}\eta}{2}(a+a^{\dagger})\Big{[}g_1 |3\rangle \langle 1|-g_1|1\rangle \langle 3|-g_2|3\rangle \langle 2|+g_2|2\rangle \langle 3|\Big{]}.
\end{equation}
The perturbed eigenstates right at the avoided crossing ($\lambda=1$) are denoted by $|\chi_{\pm}\rangle$ with corresponding eigenenergies $E_{\pm}$. Employing degenerate perturbation theory to the unperturbed eigenstates $|\phi_1,n\!-\!1\rangle$ and $|\phi_2,n\rangle$
one obtains for the energy gap
\begin{equation}
 \Delta E = E_+ - E_- =2\,|\mathcal{A}|,
 \label{mr-19}
\end{equation}
where
\begin{equation}
 \mathcal{A} =
 \langle \phi_1,n-1\,|H^{(1)}_{\rm{int}}|\,\phi_2,n\rangle.
 \label{mr-20}
\end{equation}
Under the condition $\Delta \gg g_1,g_2$ this matrix element may be approximated by
\begin{equation}
 \mathcal{A} \approx -{\rm{i}}\eta\sqrt{n}\frac{g_1g_2}{2\Delta}.
 \label{mr-21}
\end{equation}
With an appropriate choice for the phase of the unperturbed states the eigenstates $|\chi_{\pm}\rangle$ can be written as
\begin{equation} \label{mr-22}
 |\chi_{\pm}\rangle = \frac{1}{\sqrt{2}}\,\big(|\phi_1,n\!-\!1\rangle \pm\,|\phi_2,n\rangle\big).
\end{equation}

\subsubsection{Build-up of entanglement}\label{build-up}

The key features of the dynamical generation of entanglement between the electronic and vibrational degrees of freedom can
be understood with the help of a simple approximation based on the Landau-Zener picture introduced above. The general solution of the Schr\"odinger equation
\begin{equation}
 \frac{d}{dt}|\Psi(t)\rangle = -{\rm{i}} H^{\eta}|\Psi(t)\rangle
\end{equation}
in the space spanned by the Landau-Zener pair of states $|\chi_{\pm}\rangle$ is given by
\begin{equation}
\label{mr-24}
 |\Psi(t)\rangle = \alpha_+\,\rm{e}^{-\rm{i} E_+ }|\chi_+\rangle+\alpha_-\,\rm{e}^{-\rm{i} E_- t }|\chi_-\rangle.
\end{equation}
For the initial condition $|\Psi(t\!=\!0)\rangle = |\phi_2,n\rangle$, for example, we have $\alpha_{\pm}=\pm 1/\sqrt{2}$ which leads to
\begin{eqnarray}
|\Psi(t)\rangle&=&\frac{1}{2}\big{[}\textrm{e}^{-\textrm{i}E_+t}-\textrm{e}^{-\textrm{i}E_-t}\big{]}\, |\phi_1,n\!-\!1\rangle \nonumber \\
&+&\frac{1}{2}\big{[}\textrm{e}^{-\textrm{i}E_+t}+\textrm{e}^{-\textrm{i}E_-t}\big{]}\, |\phi_2,n\rangle.
\label{mr-26}
\end{eqnarray}
The right-hand side represents the Schmidt decomposition of the wave function with Schmidt coefficients
\begin{eqnarray}
 \alpha_1 &=& \frac{1}{2}\left|\textrm{e}^{-\textrm{i}E_+t}+\textrm{e}^{-\textrm{i}E_-t}\right|,
 \nonumber \\
 \alpha_2 &=& \frac{1}{2}\left|\textrm{e}^{-\textrm{i}E_+t}-\textrm{e}^{-\textrm{i}E_-t}\right|.
 \label{mr-27a}
\end{eqnarray}
The negativity of the state (\ref{mr-26}) is given by the product of the Schmidt coefficients \cite{BENGTSSON} and, hence, we obtain for the negativity of an isolated Landau-Zener state pair
\begin{equation} \label{mr-27}
 \mathcal{N}_{\textrm{LZ}}(t) = \alpha_1\alpha_2
 = \frac{1}{2}|\sin (\Delta E \, t)|.
\end{equation}
Of course, the same result is obtained for the initial state $|\Psi(t\!=\!0)\rangle = |\phi_1,n\!-\!1\rangle$. Thus, we see that within the two-state Landau-Zener approximation the entanglement dynamics at the avoided crossing is determined completely by the spectral gap $\Delta E$, leading to a periodic oscillation of entanglement between zero and $\frac{1}{2}$ with the period $T_{\rm ent}=\pi/\Delta E$. It should be noted that considering only a single isolated state pair is a valid approximation only as long as the splitting $\Delta E$ is much smaller than the separation $\omega$ between different Landau-Zener state pairs. According to Eqs.~(\ref{mr-19}) and (\ref{mr-21}) this condition sets an upper bound for the Lamb-Dicke parameter $\eta$ and for the vibrational quantum number $n$ when considering multiple Landau-Zener state pairs.

\subsubsection{Decay of entanglement}\label{entanglement-decay}

Up to now we have discussed the case $\Gamma=0$. There is a small contribution by the excited electronic state $|3\rangle$ in the dressed state $|\phi_1\rangle$ which according to Eq.~(\ref{mr-14}) is of magnitude $\sqrt{\omega/\Omega}\approx\sqrt{\omega/\Delta}$ and opens the decay channel due to spontaneous emissions.
In order to account for the effect of spontaneous emissions within the Landau-Zener model we consider the Hamiltonian
\begin{equation} \label{mr-28}
 H^0_{\rm{eff}} = H^0 - \rm{i}\frac{\Gamma}{2}\,|3\rangle \langle 3|.
\end{equation}
This is the non-Hermitian effective Hamiltonian corresponding to the master equation (\ref{mr-1}) with the Lindblad dissipator (\ref{mr-4}) which describes the coherent dynamics of the Hamiltonian $H^0$ including the finite decay widths of the dressed states caused by spontaneous emission processes. By use of this effective Hamiltonian we neglect real emission processes which lead to transitions into other Landau-Zener state pairs. Thus, we neglect here the gain of populations of other state pairs which are assumed to yield only negligible contributions to the negativity. We will see later in our numerical simulations of the full master equation that this assumption leads to a good approximation for the entanglement dynamics in the early phase of the cooling process.

\begin{figure}[h]
   \includegraphics[width=0.7\columnwidth]{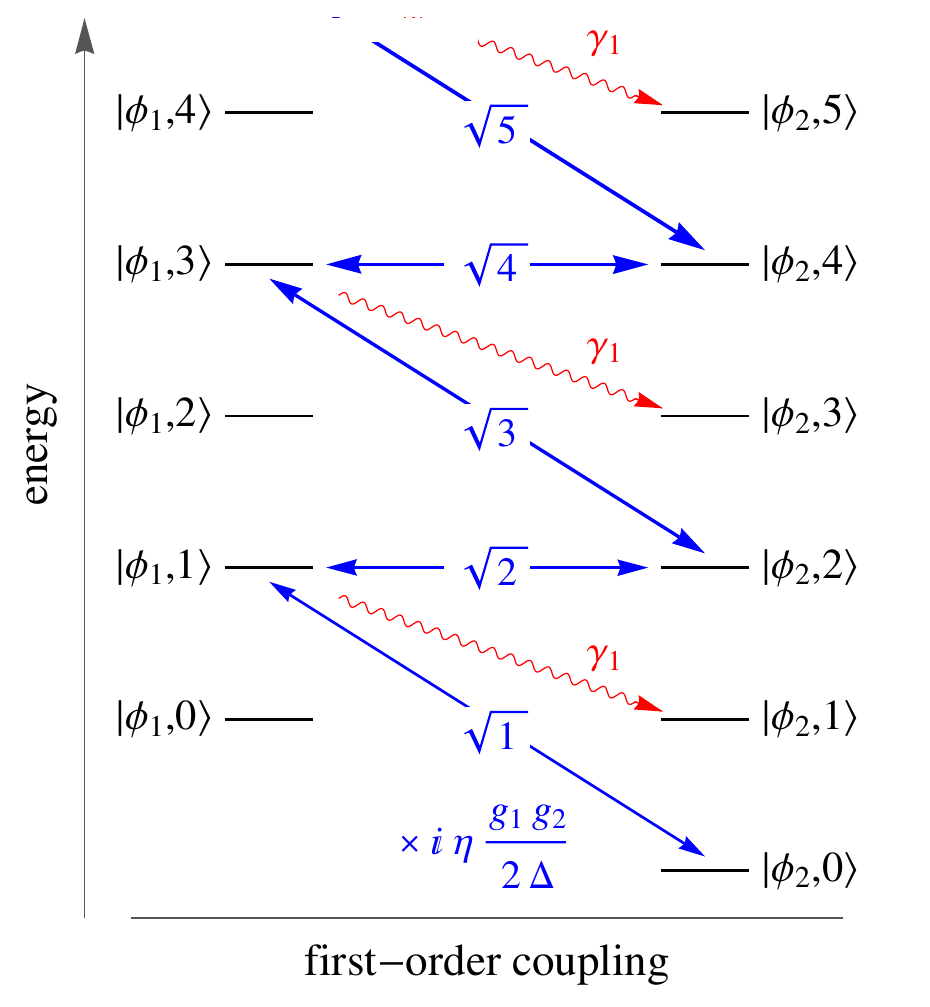}
    \caption{\label{MR-Figure-5}Elements which connect to   $|\phi_2,0 \rangle$ in first order~of~$\eta$. The dressed states are shown for $\lambda=1$. Two-photon transitions in first sidebands are marked as blue arrows, their strength  given in units of $-{\rm i}\eta g_1 g_2/(2\Delta)$, see Eq.~(\ref{mr-21}). Vibrationally diagonal transitions for spontaneous decay (red wiggly lines) occur at the rate $\gamma_1$ given in Eq.~(\ref{mr-31}).}
  \end{figure}

The complex eigenvalues of $H^0_{\rm{eff}}$ are given by $\varepsilon_{i,n}=\varepsilon_i+n\omega$, where
\begin{eqnarray*}
\varepsilon_{1}&=&\frac{1}{2}(\Delta-\textrm{i}\Gamma/2)
+\frac{1}{2}\sqrt{(\Delta+\textrm{i}\Gamma/2)^2
+g_1^2+g_2^2}, \\
\varepsilon_2&=& \Delta, \\
\varepsilon_{3}&=&\frac{1}{2}(\Delta-\textrm{i}\Gamma/2)
-\frac{1}{2}\sqrt{(\Delta+\textrm{i}\Gamma/2)^2
+g_1^2+g_2^2}.
\end{eqnarray*}
The corresponding eigenstates are of the tensor product form $|\phi_i,n\rangle=|\phi_i\rangle\otimes|n\rangle$, where
\begin{eqnarray*}
 |\phi_1\rangle &=& N_1
 \Big(g_1 |1\rangle +g_2|2\rangle+2(\varepsilon_1 -\Delta)|3\rangle\Big), \\
 |\phi_2\rangle &=& N_2 \Big(
 -g_2 |1\rangle + g_1 |2\rangle \Big), \\
 |\phi_3\rangle &=& N_3
 \Big(g_1 |1\rangle +g_2|2\rangle+2(\varepsilon_3 -\Delta)|3\rangle\Big),
\end{eqnarray*}
with the normalization factors
\begin{equation}
 N_i = \frac{1}{\sqrt{g_1^2+g_2^2+4|\varepsilon_i-\Delta|^2}}.
\end{equation}
The imaginary parts of the complex eigenvalues $\varepsilon_i$ describe the widths of the respective eigenstates. No width appears for the state $|\phi_2\rangle$. Under the condition $\Delta \gg g_1,g_2$ one finds
\begin{equation} \label{mr-30}
 \varepsilon_{1} = \Delta+\frac{\Delta}{4}\frac{g_1^2+g_2^2}{\Delta^2+(\Gamma/{2})^2}
 - {\textrm{i}}\frac{\gamma_1}{2},
\end{equation}
with the width of the state $|\phi_1\rangle$ given by
\begin{equation}
\label{mr-31}
 \gamma_1 \approx \frac{\Gamma}{4}
 \frac{g_1^2+g_2^2}{\Delta^2+(\Gamma/2)^2}.
\end{equation}
We note that the width $\gamma_1$ is much smaller than the width of the excited state $|\phi_3\rangle$ which is approximately equal to $\Gamma$. The small decay path which is active here is marked by the red wiggly lines in Fig.~\ref{MR-Figure-5}.

The dominant width is carried by the eigenstate $|\phi_3\rangle$.
This state is well separated in energy from the dressed states $|\phi_1\rangle$ and $|\phi_2\rangle$. Thus, we can derive again an approximation for the negativity by use of degenerate perturbation theory involving the states $|\phi_1,n\!-\!1\rangle$ and $|\phi_2,n\rangle$. A straightforward calculation analogous to the one carried out in Sec.~\ref{build-up} then yields the following expression for the negativity in the presence of spontaneous emission,
\begin{equation} \label{mr-32}
 \mathcal{N}_{\textrm{LZ}}(t) = \frac{\Delta E}{2}\textrm{e}^{-\gamma_1 t/2}
 \left|\frac{\sin\nu t}{\nu} +
 \frac{\gamma_1}{2\nu^2}(1-\cos\nu t)\right|,
\end{equation}
where $\nu=\sqrt{\Delta E^2-\gamma_1^2/4}$. According to this formula the entanglement dynamics at an avoided crossing is determined by the spectral gap $\Delta E$ and by the width $\gamma_1$ of the dressed state $|\phi_1,n\rangle$. Note that in the limit $\Gamma \rightarrow 0$ which implies $\gamma_1 \rightarrow 0$ the expression (\ref{mr-32}) reduces to the expression of Eq.~(\ref{mr-27}).

\section{Numerical simulation results}\label{numerics}
In this section we present results of numerical solutions of the full master equation (\ref{mr-1}) and discuss and interpret these solutions on the basis of the Landau-Zener model developed in Sec.~\ref{LZ-Model}. We shall demonstrate that the Landau-Zener picture is in fact very useful in explaining and understanding
the essential features of the complex entanglement dynamics observed in Fig.~\ref{MR-Figure-2}b. In this figure we
see a rapid build-up of entanglement, a subsequent exponentially
damped oscillation, followed by a precipitous drop and a power law
behavior and, finally, a rebirth of entanglement towards the
non-equilibrium stationary state. In the following we discuss
these features and establish a connection to the Landau-Zener model.

\subsection{Build-up of entanglement}
\begin{figure}[h]
\label{nin4}
\includegraphics [width=1\columnwidth]{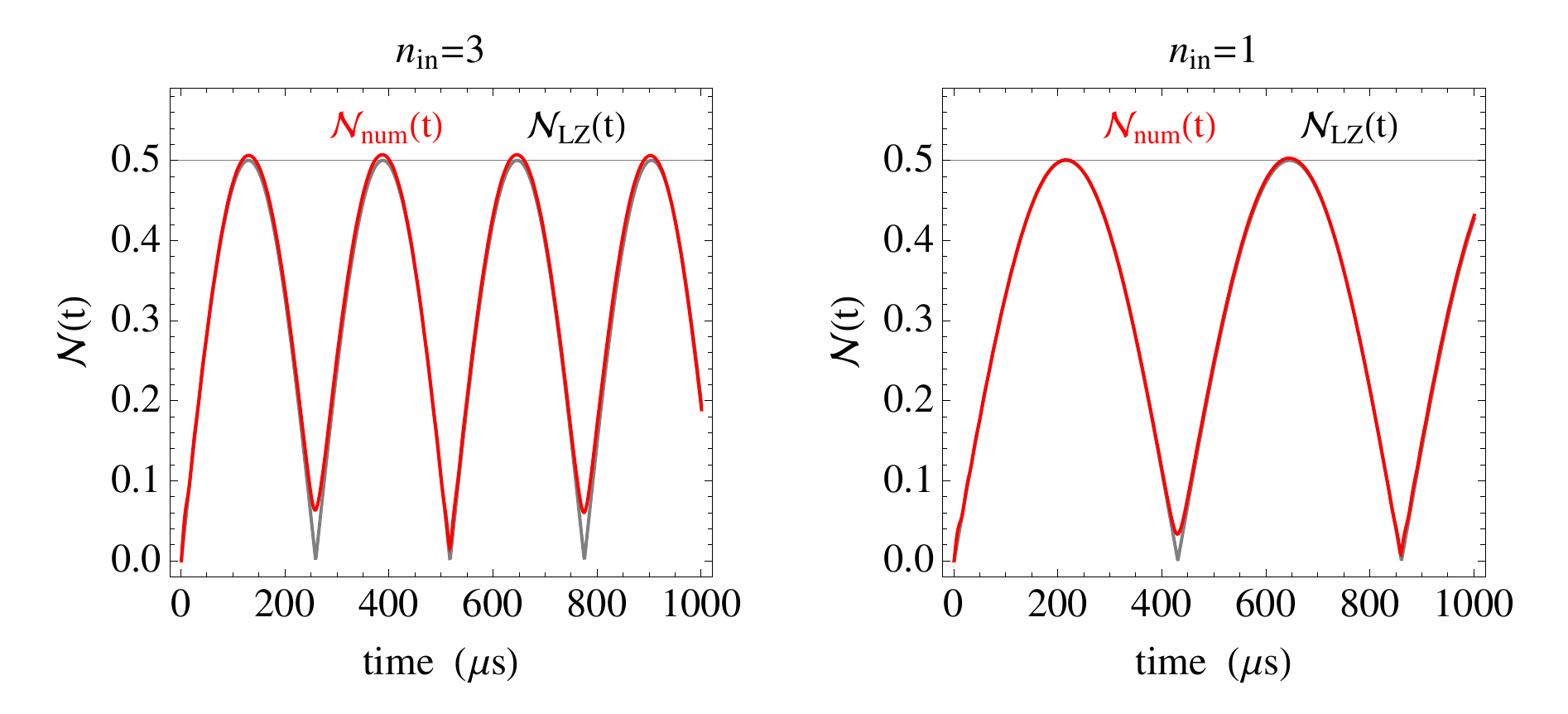}
\caption{\label{MR-Figure-6} The red lines represent the negativity based on the full numerical simulation, $\eta\!=\!0.1$. Gray lines are  the prediction~(\ref{mr-27}) based on the Landau-Zener model, using $\Delta E\!=\!0.012$ and $0.073$ respectively. Parameters are in units of $2\pi$~MHz: $\omega\!=\!0.03$, $g_1\!=\!10 g_2\!=\!1.34$, $\Delta\!=\!15$, $\Gamma\!=\!0$. }
\end{figure}

When EIT cooling begins from a state $|\phi_2,n_{\rm in}\rangle$ with $n_{\rm in}$ the initial vibrational quantum number, the two-state model predicts an initial growth of negativity from zero and subsequent oscillations according to Eq.~(\ref{mr-27})~and ~(\ref{mr-32}). This is borne out by the full simulation and particularly apparent for low values of $\Gamma$, see  Fig.~\ref{MR-Figure-2}. The full simulation also predicts this oscillating negativity when $\Gamma=0$ and in Fig.~\ref{MR-Figure-6}a and \ref{MR-Figure-6}b we compare predictions of Eq.~(\ref{mr-27}) with the results of such full simulations. The two-photon Raman transitions which connect the horizontal Landau-Zener pairs in Fig.~\ref{MR-Figure-5} are the origin for this oscillation. To a much weaker extent the two-photon Raman transitions also connect to the next-neighbor Landau-Zener pairs, above and below, see Fig.~\ref{MR-Figure-5}. This connection is not part of the two-state model, thus the small discrepancies between the full simulation and the two-state model in Fig.~\ref{MR-Figure-6} are not surprising.

Figure \ref{MR-Figure-6}a and b give results for two different initial vibrational levels, the different periodicity reflecting the $n$-dependence in Eq.~(\ref{mr-21}). The negativity oscillation period is $T_{\rm ent}=\pi/\Delta E$. Since $\Delta E \propto \sqrt{n_{\textrm{in}}}$, the choice of different $n_{\textrm{in}}$ values leads to different oscillation periods. Note that the full simulation predicts negativities which slightly exceed the value of 1/2, the maximal value possible for a tensor product space in which one factor is two-dimensional. The reason is that the space describing the internal degree of freedom of the atom is three-dimensional here, the maximal value for the negativity being thus equal to 1. However under EIT cooling conditions the excited electronic state is extremely rarely occupied and the actual negativity therefore stays far away from this maximal value.

\subsection{Decay of entanglement}
When taking into account the small contamination of $|\phi_1\rangle$ by excited state character and its decay width $\gamma_1$, we expect decoherence to occur, its magnitude being weighted by the amplitude of $|\phi_1\rangle$  in $|\Psi(t)\rangle$. In the two-state model the trapped atom is initially in $|\phi_2\rangle$ occupying the vibrational quantum number $|n_{\textrm{in}}\rangle$ with $\mathcal{N}=0$.  When the lasers are turned on with $\lambda=1$, the atom oscillates between states $|\phi_2,n_{\textrm{in}}\rangle$ and $|\phi_1,n_{\textrm{in}}\!-\!1\rangle$, the negativity climbing periodically to a value 1/2. When during this cycle the atom finds itself in the state $|\phi_1,n_{\textrm{in}}\!-\!1\rangle$, it may suffer relaxation by spontaneous emission to each state of the lower Landau-Zener pair, $|\phi_2,n_{\textrm{in}}\!-\!1\rangle$ and $|\phi_1,n_{\textrm{in}}\!-\!2\rangle$. Both contain a fraction of the electronic ground states $|1\rangle$ and $|2\rangle$, but in the Lamb-Dicke limit the spontaneous transition to $|\phi_2,n_{\textrm{in}}\!-\!1\rangle$, indicated by the red wiggly arrows in Fig.~\ref{MR-Figure-5}, is stronger by a factor $\propto \eta^{-2}$. This loss is accounted for within the Landau-Zener model by the calculations presented in Sec.~\ref{entanglement-decay}.

\begin{figure}[h]
\label{Negativcompare100}
\includegraphics [width=1\columnwidth]{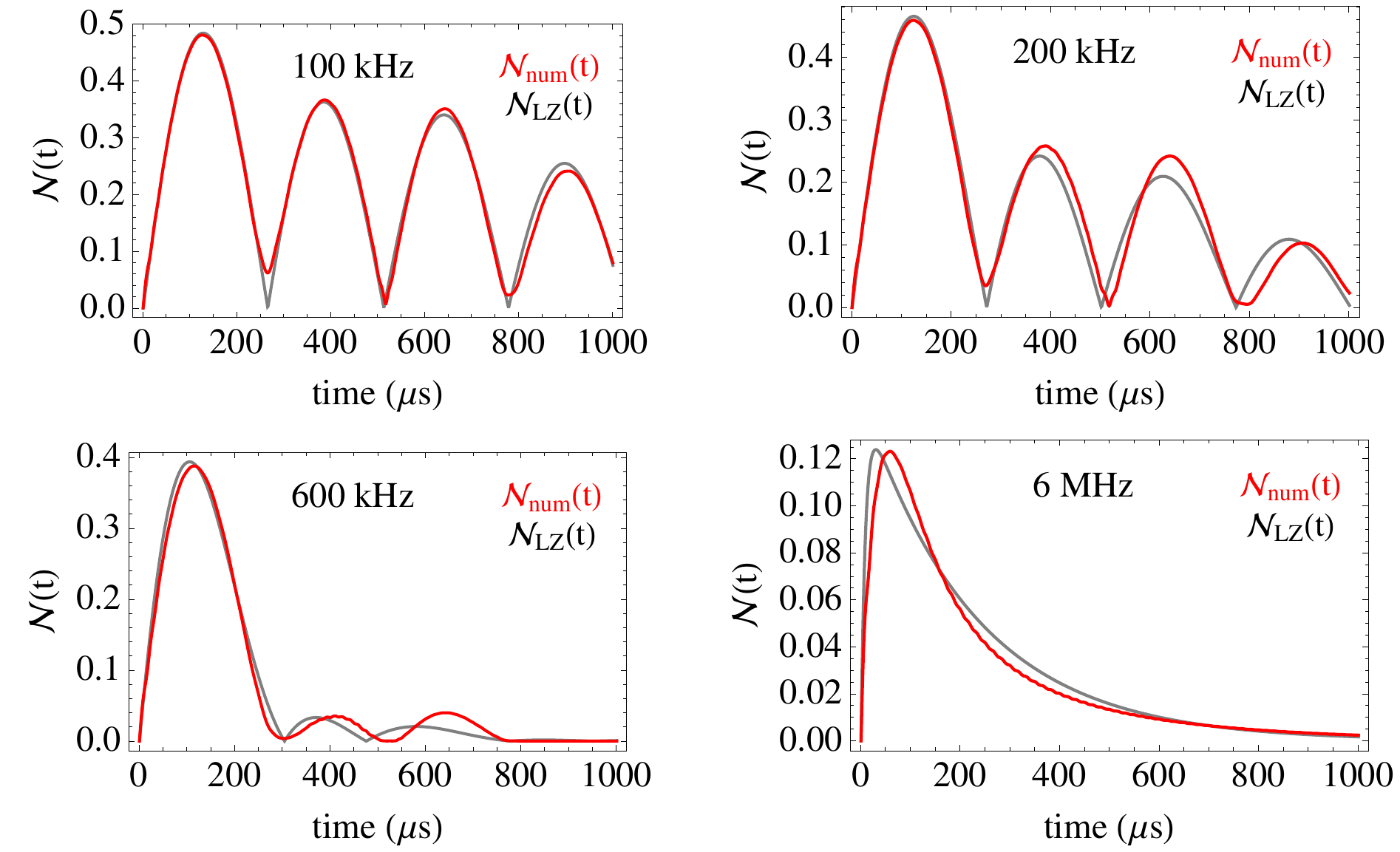}
\caption{\label{MR-Figure-7}Transition from underdamped to overdamped oscillatory behavior of the negativity for various values of $\Gamma$, $\eta\!=\!0.1$, $n_{\rm{in}}\!=\!3$. 
Fits of Eq.~\eqref{mr-32} are shown by the gray lines (Table~\ref{fit}). Red lines show the negativity from the full simulation.}
\end{figure}

In Fig.~\ref{MR-Figure-7} we compare these predictions of the two state model with results from a full simulation at various values of $\Gamma$. In this comparison we have used the quantities $\Delta E$ and $\gamma_1$ as fit parameters.
\begin{table}[h]
\caption{Fitted parameters used in Fig.~\ref{MR-Figure-7} in comparison to the predictions from Eq.~(\ref{mr-19}) and~(\ref{mr-32}).}
\centering
  \begin{tabular}{r| c c l l l }
      $ \Gamma \quad$ & $\,\, \,  \Delta E$({fit})&$ \,\, \, \Delta E$ \,\, \,  &$\, \, \gamma_{1}$(fit)\, \,&$\, \, \, \gamma_1$   \\
      \hline  \hline
   $100\, \textrm{kHz}$& 0.0126 & 0.015& 0.0016& 0.00063 \\ 
   $200\, \textrm{kHz}$& 0.0126 & 0.015& 0.0016& 0.00126 \\ 
   $600\, \textrm{kHz}$& 0.0146 & 0.015& 0.0062& 0.00377 \\ 
   $6\, \textrm{MHz}$& 0.0316 & 0.015& 0.1141& 0.036 \\ 
  \end{tabular}
 \label{fit}
 \end{table}
We see that the behavior observed for the negativity mimics that of a damped oscillator, showing a transition from underdamped to overdamped motion. As is illustrated in Fig.~\ref{MR-Figure-7} the simple formula \eqref{mr-32} provides an excellent approximation of the entanglement dynamics for $\gamma_1/2 < \Delta E_n$, and even for intermediate values of the spontaneous emission rate the qualitative behavior of the negativity is reproduced.
It is remarkable that Eq.~\eqref{mr-32} correctly describes the transition from the underdamped to the overdamped motion of the negativity at $\gamma_1/2=\Delta E$, compare Figs.~\ref{MR-Figure-7}c and \ref{MR-Figure-7}d. Using the approximations $\Delta E\approx\eta\sqrt{n}g_1g_2/\Delta$ and
$\gamma_1 \approx \Gamma g_1^2/(4\Delta^2)$ this transition is predicted to occur at the value $\Gamma = 8\eta\sqrt{n}\Delta g_2/g_1$, which nicely fits to the results of our numerics.

The interpretation of the damping of negativity is that after a spontaneous event the system has forgotten the history of coherent oscillation in the former Landau-Zener pair and starts to rebuild  negativity in the new Landau-Zener pair, at least until the moment that it finds itself in  state $|\phi_1,n\!-\!2\rangle$ and renewed relaxation may occur. We see from this that spontaneous emission introduces a dephasing of the coherent oscillations of $\mathcal{N}(t)$ in different Landau-Zener pairs. This means that the negativity computed for each pair is uncorrelated in phase from that of it's next neighbors. The total negativity is very sensitive to decoherence, the total negativity being smaller than the sum of the negativities over all Landau-Zener pairs. This example illustrates the subadditivity character of the negativity, in particular the requirement of phase correlation in superposition of entangled state parts.
On this basis we may attempt to discuss in a broader sense the decay of entanglement seen in Fig.~\ref{MR-Figure-2}b.

\begin{figure}[h]
\label{array1}
\includegraphics [width=0.9\columnwidth]{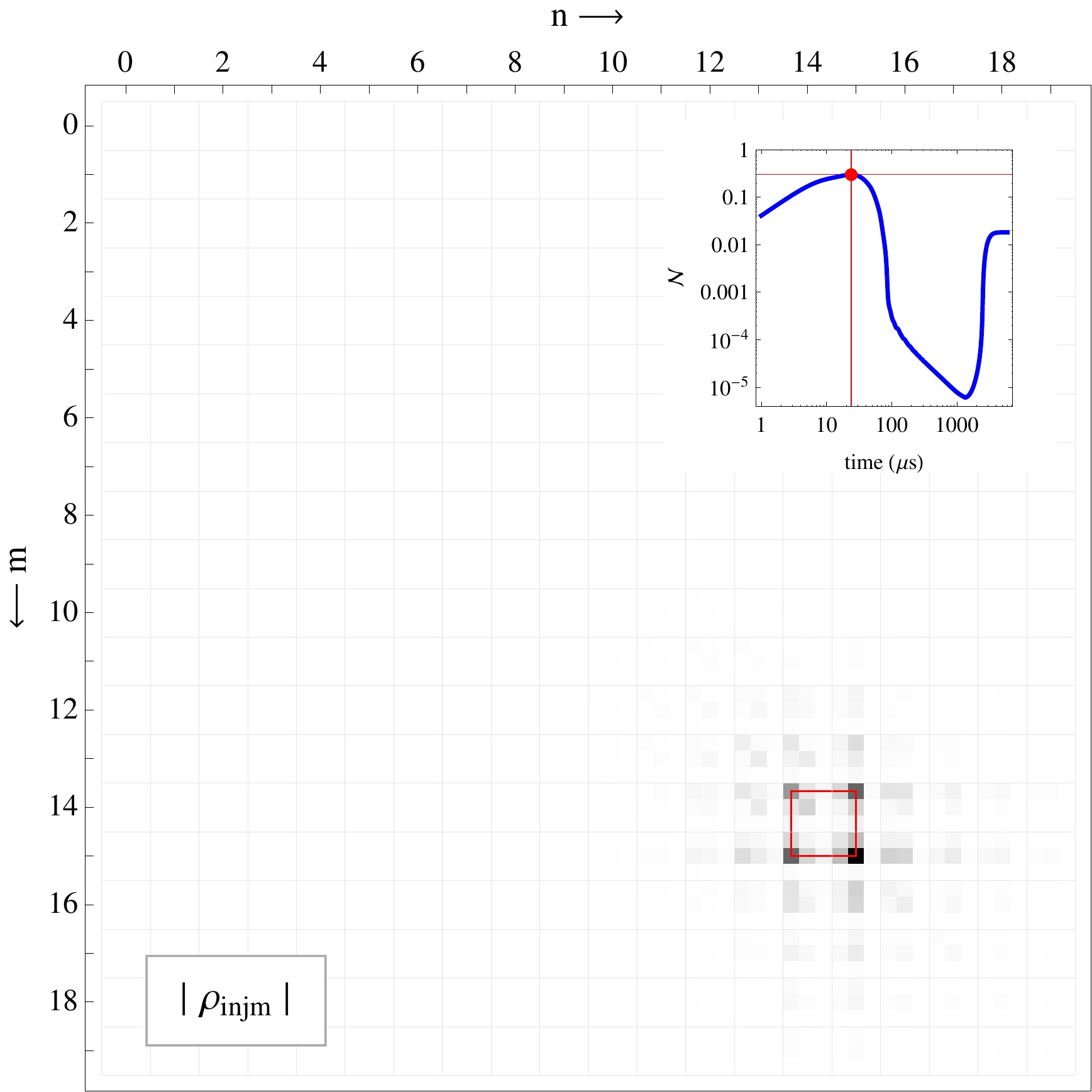}
\caption{\label{MR-Figure-8} The absolute density matrix elements are shown at a time $t=24\,\mu$s, when negativity maximizes for these parameters, see text. The Landau-Zener model for negativity is confined to the elements at the corners of a single red square.}
\end{figure}
As cooling ensues, a spread of the density matrix elements occurs.  This is evident from Fig.~\ref{MR-Figure-8} which gives a matrix representation of the absolute values of elements of the density operator. Each square representing a vibrational quantum number contains nine elements for electronic population and coherences.  The result shown here corresponds to the conditions $\Gamma\!=\!6$, $g_1\!=\!5\, g_2\!=\!1.34 $, $\omega\!=\!0.03$, $\Delta\!=\!15$~in units of $2\pi$~MHz, and $\eta\!=\!0.1$ with  $n_{\rm in}\!=\!15$.  Note that here the negativity reaches a higher value, and does so faster than in Fig.~\ref{MR-Figure-2}b, where $g_1/g_2\!=\!10$.  Reasons are the two times higher two-photon Rabi-frequency and the increased coupling (\ref{mr-21}) which scales with $\sqrt{n_{\rm in}}$. Fig.~\ref{MR-Figure-8} shows the density matrix at a time when the maximal negativity is reached (see inset). In our Landau-Zener model the entanglement stems solely from the contributions of the elements of a single Landau-Zener pair, marked by the corners of the red square in Fig.~\ref{MR-Figure-8}. The negativity based on the trace norm takes into account all the density matrix elements from the master equation and we see from Fig.~\ref{MR-Figure-8} that population and coherences have already spread outside the initial Landau-Zener pair.

\subsection{Entanglement and spontaneous emission}
\begin{figure}[h]
\includegraphics [width=0.72\columnwidth]{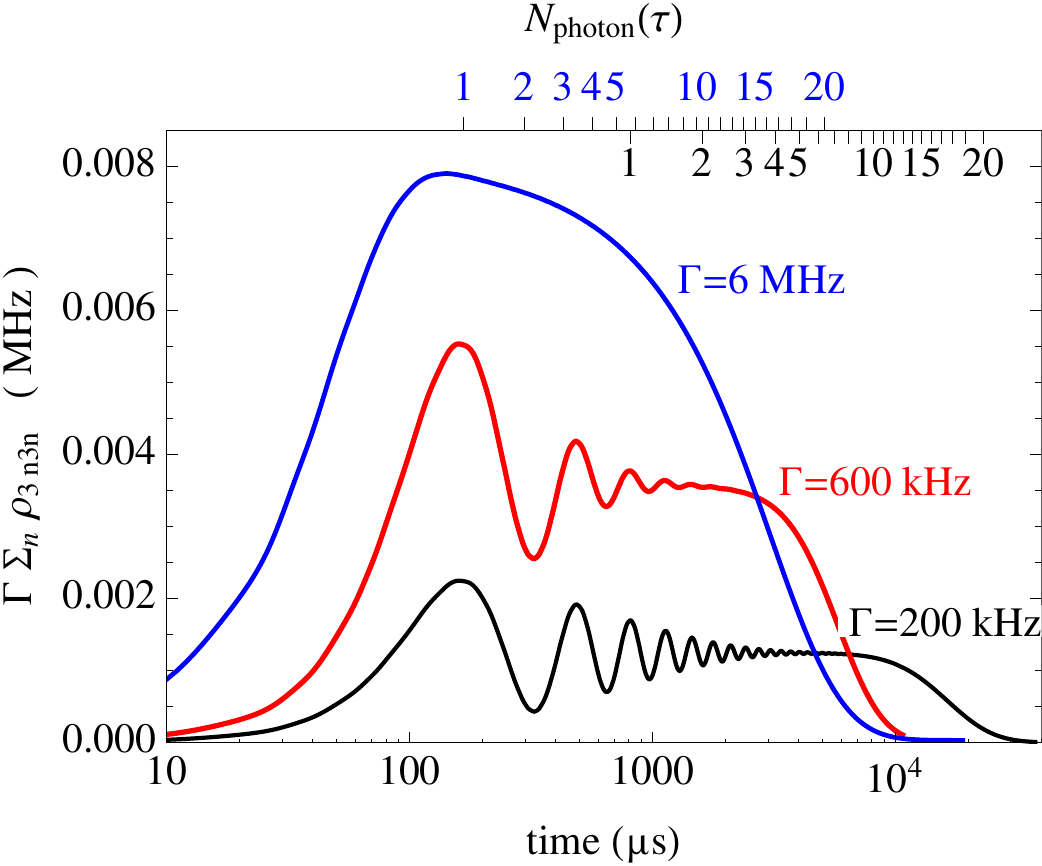}
\caption{\label{MR-Figure-9} Rate of spontaneous emission during cooling for the three cases shown in Fig~\ref{MR-Figure-2}. Integral values of the accumulated rate (the number of spontaneous photons emitted) are given by the numbers on the top axis for $\Gamma=6$ ~MHz and 200~kHz respectively.}
\end{figure}

\begin{figure}[h]
\includegraphics [width=0.72\columnwidth]{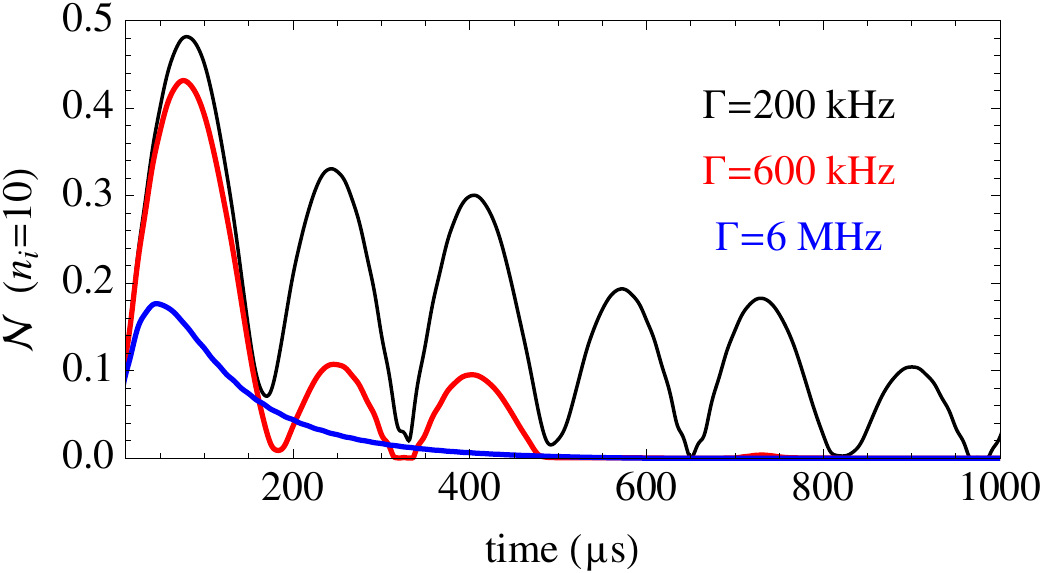}
\hspace{1mm}
\includegraphics [width=0.72\columnwidth]{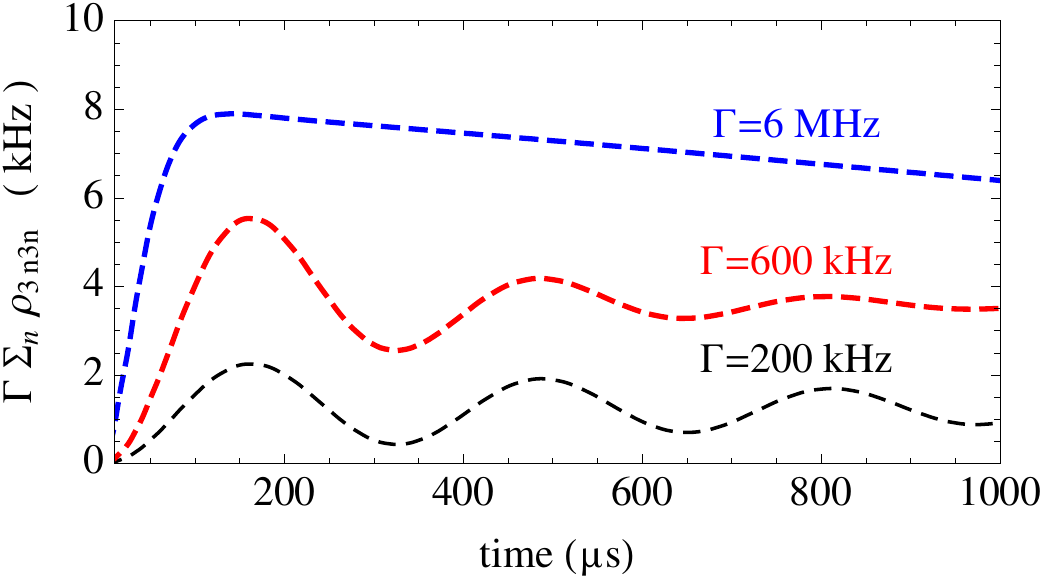}
\caption{\label{MR-Figure-10}The rate of spontaneous emission (lower figure) is out of phase with the oscillations in negativity (upper figure) and occurs preferentially at every second minimum of negativity, when the state $|\phi_1,n-1\rangle$ is most strongly populated.}
\end{figure}

Returning to the full model simulation which gives us the results shown in Fig.~\ref{MR-Figure-2}, we evaluate the rate of emission of spontaneous photons from the full density matrix using
\begin{equation}
\label{mr-33}
R(t)=\Gamma\sum_n \rho_{3n3n}(t)\, .
\end{equation}
Figure~\ref{MR-Figure-9} shows the development of this rate during the cooling process for the three cases considered in Fig.~\ref{MR-Figure-2}. From Eq.~(\ref{mr-33}) we may estimate the  total number of spontaneously emitted photons (approximately 2 spontaneous photons per vibrational level cooled \cite{RBH}) and also the time of emission. We may argue that at times $\tau$ when
\begin{equation}
\label{mr-34}
\textrm{N}_{\textrm{photon}}(\tau)=\Gamma \int_0^\tau\,dt\,\sum_n \rho_{3n3n}(t)
\end{equation}
climbs to the next integer value a spontaneous photon has left the trapped atom.
The sequence of integer values $\textrm{N}_{\textrm{photon}}(\tau)$ is marked along the top axis in Fig.~\ref{MR-Figure-9} for the cases $\Gamma=6$ and 0.2 MHz. We see that spontaneous emission practically ceases when the cooling limit is reached. The connection between the oscillatory nature of the rate of emission of spontaneous photons and the rate of oscillations of the negativity in Fig.~\ref{MR-Figure-2}b can be understood by comparing these results directly, see Figure \ref{MR-Figure-10}. The rate of spontaneous emission is perfectly out of phase with negativity and maximizes whenever the state $|\phi_1,n\!-\!1\rangle$ is most strongly populated, that is at every second minimum in negativity, in perfect agreement with the prediction of the Landau-Zener model.

\begin{figure}[h]
\label{array2}
\includegraphics [width=0.9\columnwidth]{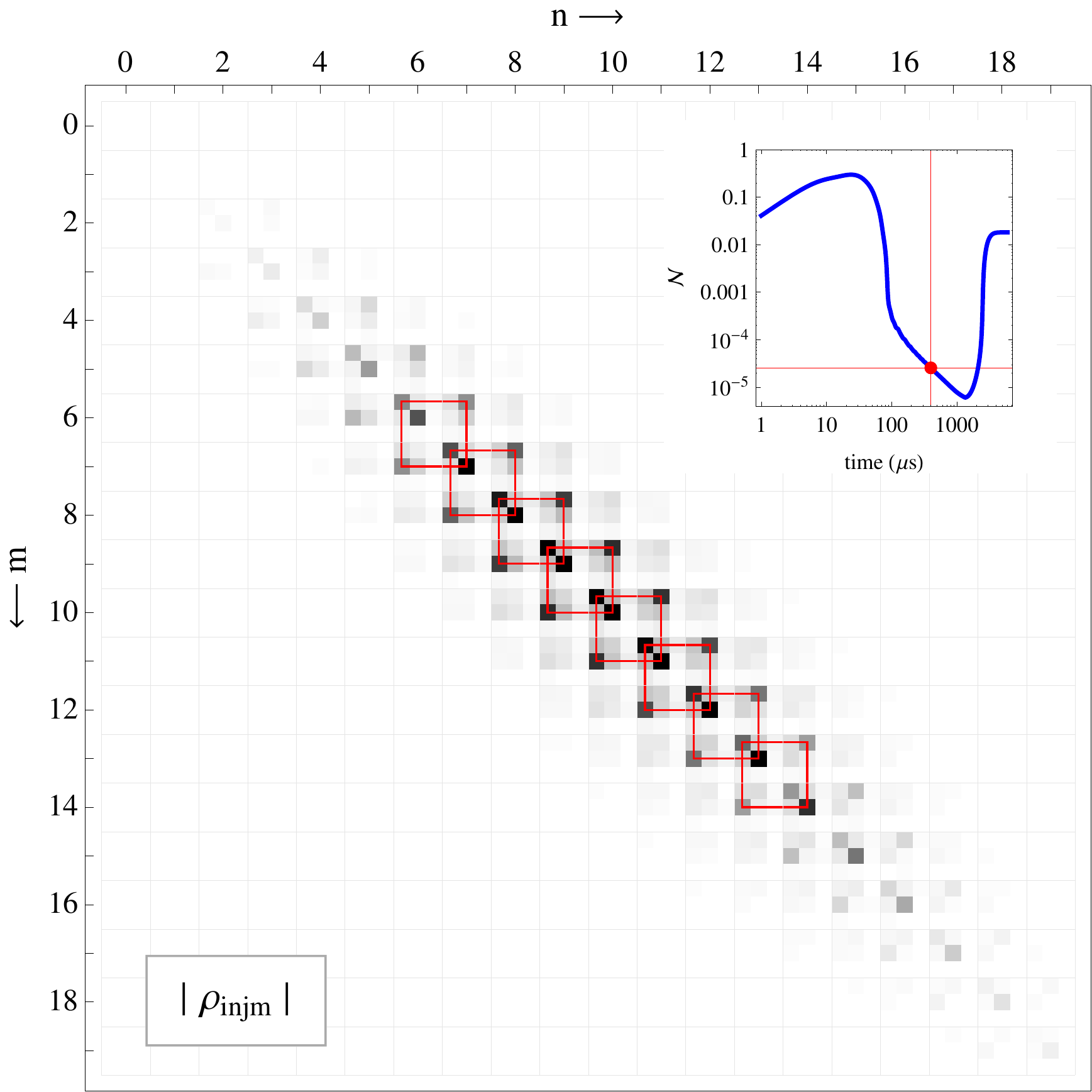}
\caption{\label{MR-Figure-11} The absolute density matrix elements are shown at a time $t=400\,\mu$s, when negativity has dropped to near zero. Parameter as in Fig.~\ref{MR-Figure-8}. Populations and coherences have by now spread over a wide domain of the density matrix. The action of many neighboring Landau-Zener pairs is clearly evident as marked by the red squares.}
\end{figure}

To explore this situation at later times we show in Fig.~\ref{MR-Figure-11} the distribution of density matrix elements at a time when the negativity is near its lowest values. Again, red squares mark the elements of individual Landau-Zener pairs in the density matrix. It is evident that following the initial climb to a high negativity a drop into a region of low negativity ensues, which is caused by spontaneous emission and which reflects the incoherent distribution of density over many elements. Nevertheless, substantial coherences appears within each square, as is evident by the strength of the elements at the off-diagonal corners. Incoherency is a consequence of the statistical character of spontaneous emission. Only it leads to a transfer of population from one square to it's lower neighbor and, hence, to cooling.

\subsection{Rebirth of entanglement}
\begin{figure}[h]
\label{fidel}
\includegraphics [width=0.48\columnwidth]{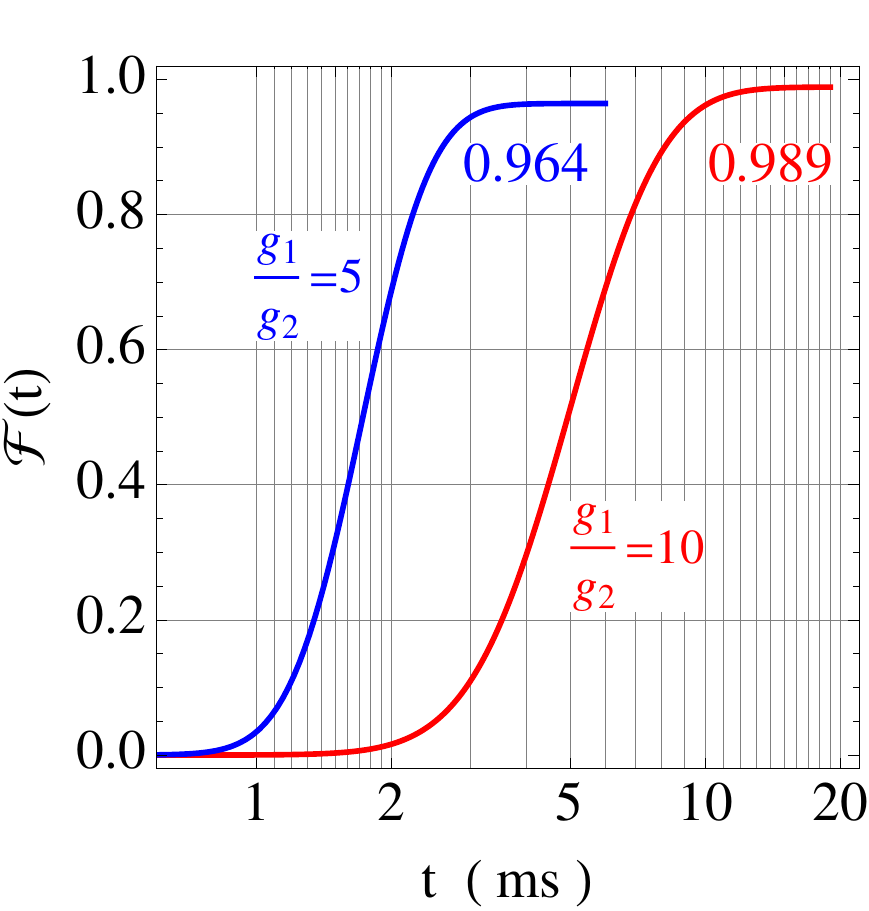}
\hspace{1mm}
\label{array3}
\includegraphics [width=0.44\columnwidth]{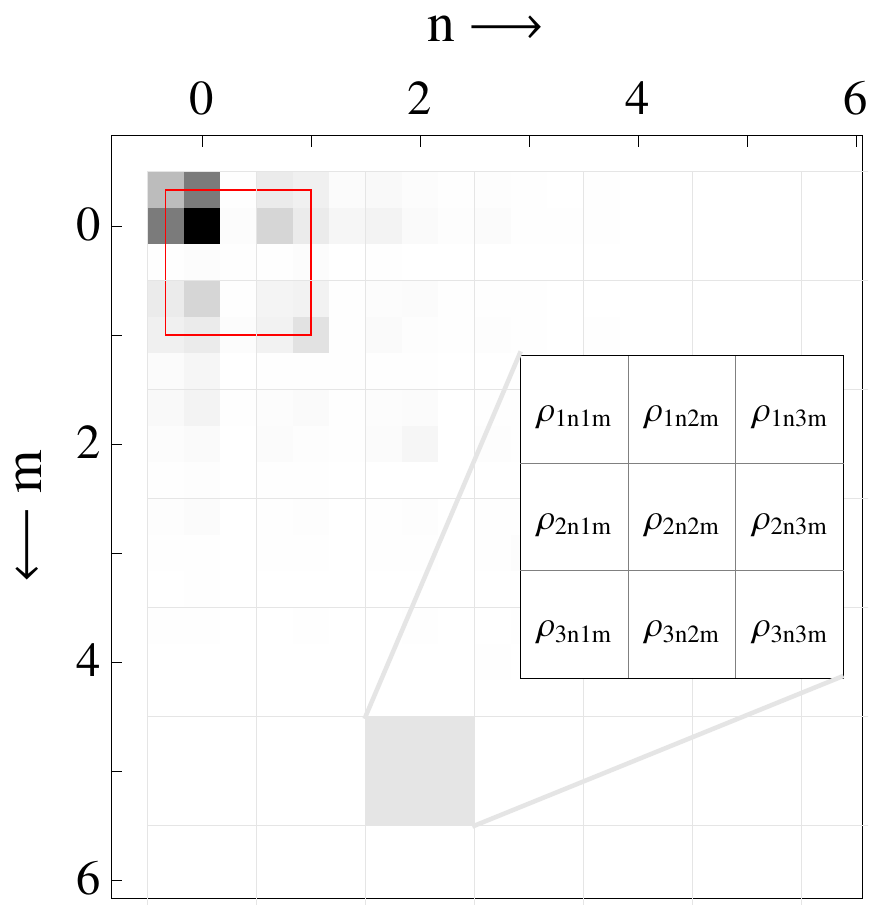}
\caption{\label{MR-Figure-12} (a) The fidelity of the dark state $|\phi_2,0\rangle$ as a function of time at $\Gamma=6$~MHz for the parameters of Figs.~\ref{MR-Figure-2} and \ref{MR-Figure-11}, respectively. (b) Matrix representation of the cooling limit ($t=6000\, \mu$s) for the conditions depicted in Figs.~\ref{MR-Figure-8} and \ref{MR-Figure-11}.}
\end{figure}

During the cooling process the atom is driven into the unique stationary state $\rho_{\rm stat}$ of the master equation (\ref{mr-1}). The final value ${\mathcal{N}}(\infty)$ of the negativity represents the entanglement of this state which has a dominant contribution from $|\phi_2,0\rangle$, the product of the dark state and the vibrational ground state of the trap with zero negativity. However, $\rho_{\rm stat}$ contains a small amount of entanglement which is controlled by the off-diagonal transitions from the dark state. This can be seen from the dynamics of the fidelity \cite{Chuang2} of the dark state,
\begin{equation}
 \mathcal{F}(t) = \langle \phi_2,0| \rho(t)|\phi_2,0\rangle,
\end{equation}
which is shown in Fig.~\ref{MR-Figure-12}a. We find that the fidelity at the end of the cooling cycle (the stationary state) is approximately equal to $0.989$ for $g_1/g_2=10$ and $0.964$ for $g_1/g_2=5$ for the parameters used in Fig.~\ref{MR-Figure-2} with $\Gamma=6$~MHz.

The observed revival of the negativity at the end of the cooling process thus reflects the amount of entanglement in the final stationary state. To estimate the final negativity for small values of $\Gamma$ we determine the interacting ground state $|\chi_0\rangle$ of the Hamiltonian $H^{\eta}$ to first order in the Lamb-Dicke parameter. Taking into account the coupling of the state $|\phi_2,0\rangle$ to the excited state $|\phi_1,1\rangle$ (see Fig.~\ref{MR-Figure-5}) through the matrix element
\begin{equation}
 {\mathcal{A}} = \langle \phi_1,1|H_{\rm{int}}^{(1)}|\phi_2,0\rangle \approx
 -{\rm{i}}\eta \frac{g_1 g_2}{2\Delta},
 \label{mr-45}
\end{equation}
we find the ground state (normalized to first order in $\eta$)
\begin{equation}
 |\chi_0\rangle \approx |\phi_2,0\rangle
 - \frac{{\mathcal{A}}}{2\omega} |\phi_1,1\rangle \, .
\end{equation}
The negativity of this state is given by
\begin{equation} \label{mr-38}
 {\mathcal{N}}(|\chi_0\rangle) = \frac{|{\mathcal{A}}|}{2\omega} = \eta \frac{g_1g_2}{4\omega\Delta} \, .
\end{equation}
In the limit $\Gamma \to 0$ this negativity should coincide with
the value ${\mathcal{N}}(\infty)$ for the stationary state negativity. In fact, this prediction is in quantitative agreement with the results for the full model in Fig.~\ref{MR-Figure-2}b: For the parameters chosen in this figure Eq.~(\ref{mr-38}) yields ${\mathcal{N}}(\infty)=0.01$ in very good agreement with the simulations results for $\Gamma=200$ and 600 kHz. For $\Gamma$=6 MHz the two-photon Rabi frequency is no longer much faster than the spontaneous rate and, hence, in this case the result of the numerical simulation in Fig.~\ref{MR-Figure-2}b falls slightly below the prediction of Eq.~(\ref{mr-38}) by about 20\%.

\subsection{Entanglement control}\label{entanglement-control}

The discussion above shows that the entanglement in the trapped atom's degrees of freedom is imprinted by two-photon transitions between the two ground states. In the absence of an external radiation field the two ground states are however free from decay.
We therefore expect that entanglement can permanently be imprinted in the trapped atom system. Here we explore this effect by switching off both lasers at a certain time during the cooling cycle. We do not consider the case that only one laser is turned off, as this will obviously lead to a complete repumping into one of the ground state levels by spontaneous emission with a steady-state negativity of zero.

\begin{figure}[h]
\includegraphics [width=0.431\columnwidth]{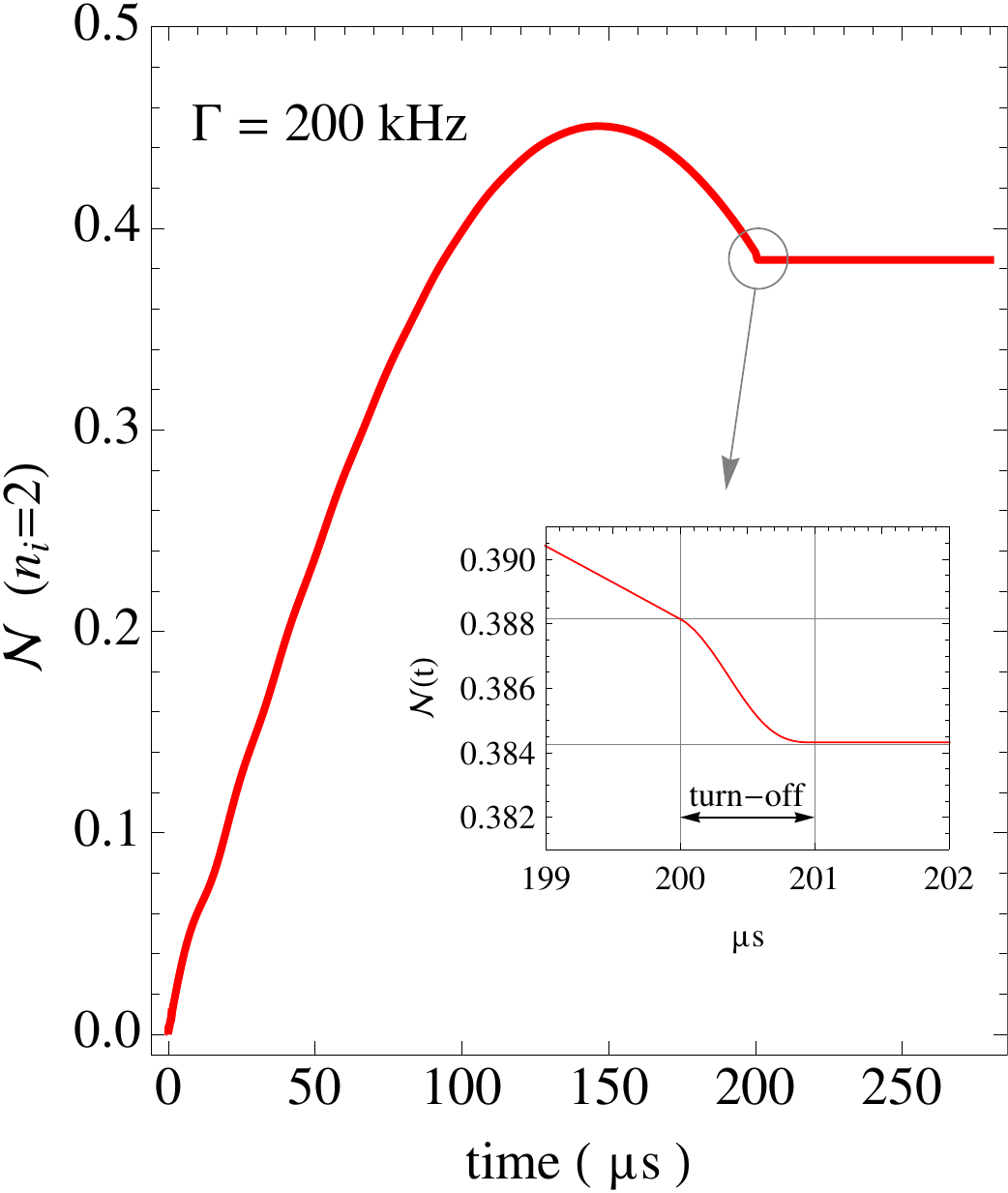}
\hspace{2mm}
\includegraphics [width=0.46\columnwidth]{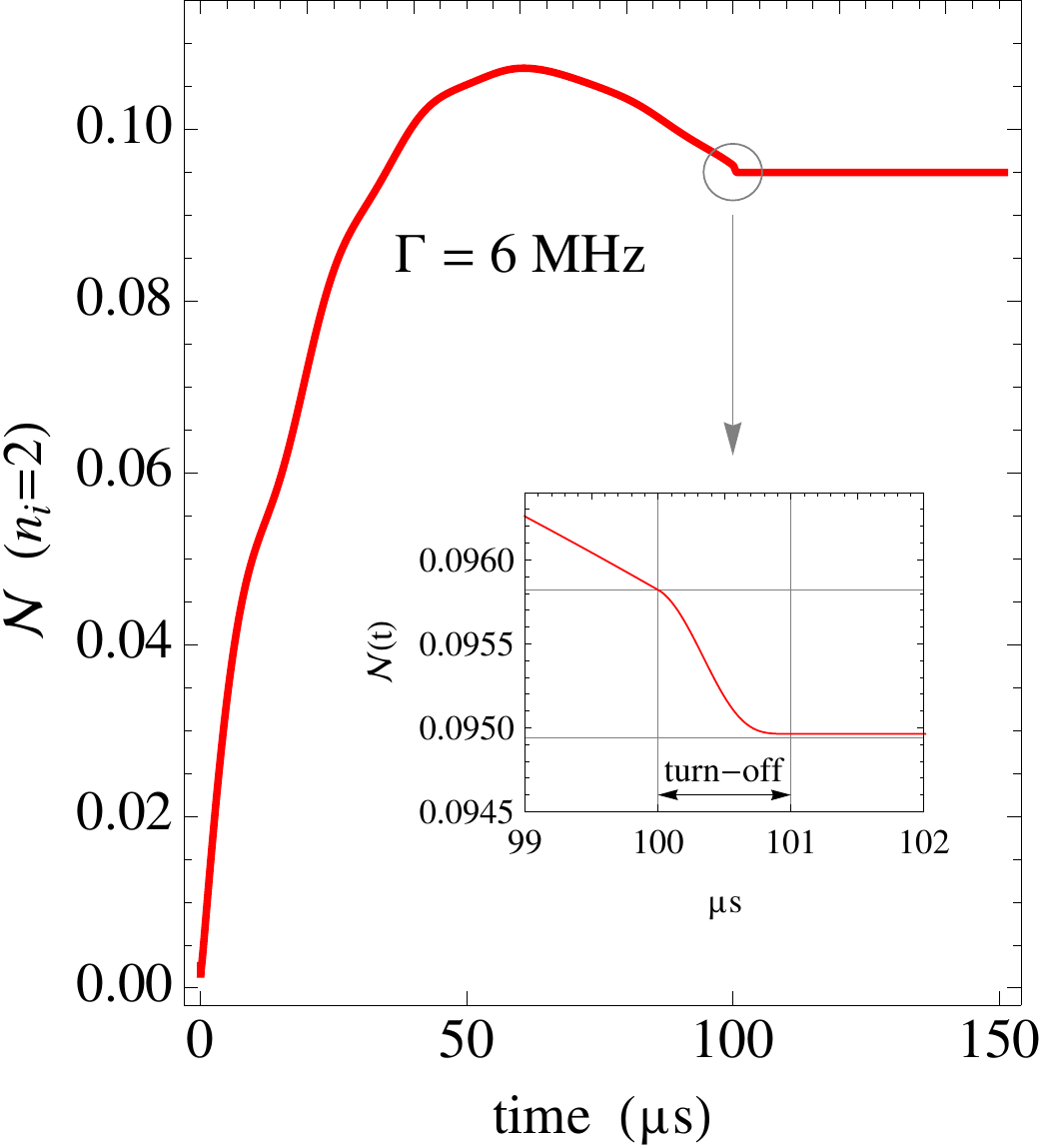}
\caption{\label{MR-Figure-13} A sudden turnoff of both lasers  
nearly preserves the negativity. The close-ups show that a minute fraction of negativity suffers decay  which stems from the tiny contribution of the excited electronic state. Parameters as in Fig.~\ref{MR-Figure-2}b.
}
\end{figure}

The turnoff of the laser fields starts at some time $t_{\rm off}$ and lasts for a time $\Delta t$, and is described by time-dependent Rabi frequencies $g_i(\lambda(t))=g_i\lambda(t)$ with the following form for the function $\lambda(t)$,
\begin{equation}
 \lambda(t) =\!\left\{
\begin{array}{ll}
 1 & {\mbox{for}} \;\; t<t_{\rm{off}}, \\
 1\!-\!\sin^2{\frac{\pi(t-t_{\rm off})}{2\Delta t}} & {\mbox{for}} \;\; t_{\rm{off}}<t<t_{\rm{off}}\!+\!\Delta t\, , \\
 0 & {\mbox{for}} \;\; t>t_{\rm{off}}+\Delta t.
\end{array}
 \right.
 \label{mr-47}
\end{equation}
To model a sudden switch off we choose $\Delta t=1\mu{\rm s}$. Two examples of simulation results for the full master equation are shown in Fig.~\ref{MR-Figure-13}. We notice that the negativity is practically frozen at its value at the turnoff time.
A closer inspection reveals a minute decrease in negativity on the time scale of spontaneous emission, which is connected to the small contribution to the negativity from the excited electronic state.

The observed freezing of negativity can be understood in general terms by means of the following argument. The Hamiltonian of the total system after switching off the lasers takes the form $H_{\rm cm} + H_{\rm el}$, describing an uncoupled time-evolution of the internal and the external degrees of freedom. Since the contribution of the excited electronic state is very small under EIT conditions, the total time-evolution operator after switch-off is given by a tensor product of local unitary operators which leaves invariant any measure for entanglement.

\begin{figure}[h]
\includegraphics [width=0.7\columnwidth]{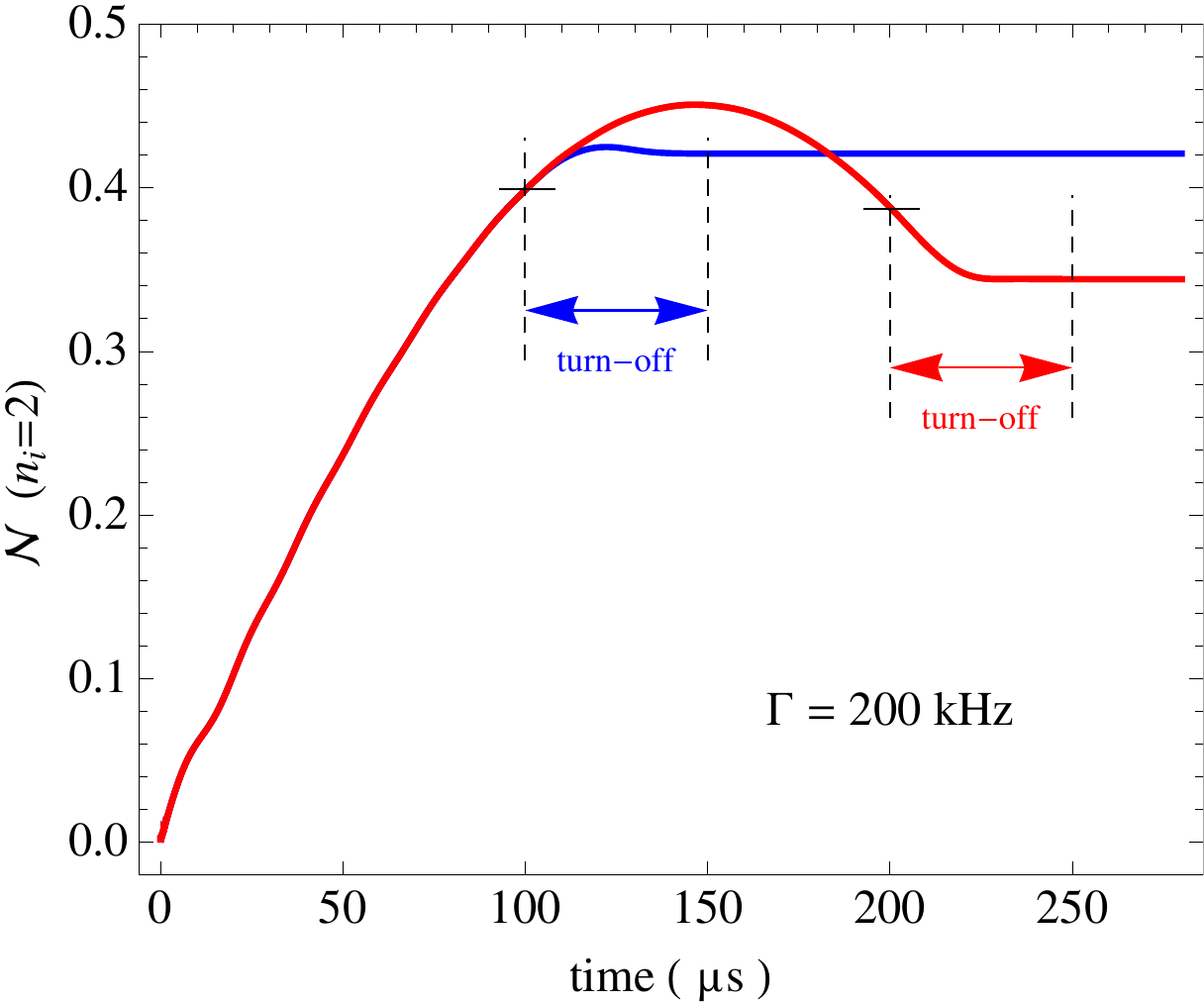}
\caption{\label{MR-Figure-14} Adiabatic turnoff of both lasers over a time period of  $\Delta t=50\,\mu{\rm s}$.  The two examples shown refer to  $t_{\rm off}=100\,\mu{\rm s}$ and $t_{\rm off}=200\,\mu{\rm s}$.}
\end{figure}

When turning off both lasers slowly a more adiabatic transition regime is observed, see Fig.~\ref{MR-Figure-14}.
Now the negativity during turnoff mirrors the decreasing Rabi frequencies in the coupling element (\ref{mr-21}), $ \mathcal{A} \approx -{\rm{i}}\eta\sqrt{n}\frac{g_1g_2}{2\Delta}$.  This element controls the oscillations of negativity with a single Landau-Zener pair.  The adiabatic decrease in the two-photon Rabi frequency diminishes  this oscillation, regardless of whether we are in the rising or decreasing regime of negativity. Similar observations were made at $\Gamma=6$~MHz, albeit at overall lower values of negativity.

\section{Discussion and conclusions}\label{conclu}

We have discussed here in detail the entanglement dynamics in EIT laser cooling of trapped atoms, employing the negativity as a measure for entanglement. We have demonstrated that essential features of the dynamical behavior of the negativity can be explained in terms of a conceptually simple model which is based on Landau-Zener state pairs at avoided crossings in the spectrum of the full atom-field Hamiltonian. As we have seen by the comparison with numerical simulations of the full master equation, the model correctly describes the entanglement oscillations and their decay in the early stage of the cooling process, as well as the revival of entanglement towards the final stationary state of the atom. While these features can be explained qualitatively and quantitatively on the basis of the Landau-Zener model, a deeper understanding of other features, like the observed power law decay of negativity and its precipitous drop in the transient regime of cooling, is still lacking. This requires a more sophisticated treatment of the dynamical transfer of negativity between different Landau-Zener state pairs induced by spontaneous emission processes. In this context it may also be of interest to explore the situation when two or more initial vibrational states are driven into  coherent oscillations of negativity.

In Sec.~\ref{entanglement-control} we have discussed a scheme for the freezing of the entanglement between internal and external atomic degrees of freedom by a rapid turnoff of the laser fields. This scheme may find further interesting applications in the entanglement control of a much more general class of physical
systems which are driven by external fields. Moreover, it seems worth studying in more detail other control methods involving an adiabatic turnon and turnoff, or suitably designed pulse shapes and sequences of pulses of the external driving fields.

Finally, the theoretical picture developed here in order to understand the dynamical behavior of the negativity points towards fundamental distinctions between the entanglement which is generated by laser cooling of atoms in the EIT scheme on one hand, and by VSCPT on the other. We briefly discuss these distinctions in the following.

According to the Landau-Zener model constructed in Sec.~\ref{LZ-Model}, the initial phase of the entanglement dynamics in EIT cooling can be understood as a periodic oscillation between the eigenstates
\begin{equation}
 |\chi_{\pm}\rangle = \frac{1}{\sqrt{2}}\big(|\phi_1,n-1\rangle \pm |\phi_2,n\rangle\big)
\end{equation}
of the trapped atom-laser field Hamiltonian $H^{\eta}$ at a Landau-Zener splitting. These states are maximally entangled and lead to a periodic oscillation of the negativity between zero and $\frac{1}{2}$ with a period $T_{\rm ent}=\pi/\Delta E$ which is determined by the energy gap $\Delta E$ at the avoided crossing.
The periodic oscillation of the negativity is interrupted stochastically by spontaneous emissions. After a spontaneous emission event a build-up of entanglement in the next lower Landau-Zener pair ensues. As this contribution is however unrelated in phase to that of the parent pair, the subadditivity of the negativity leads to an overall decrease of the amount of entanglement as cooling proceeds. The dominant final state of the system which is reached in EIT cooling is the product state $|\phi_2,0\rangle$. However, a small degree of entanglement appears in the final state of EIT cooling which is mainly due the coupling
of the states $|\phi_2,0\rangle$ and $|\phi_1,1\rangle$, as is indicated by the lowest diagonal blue arrow in Fig.~\ref{MR-Figure-5}. In the limit of $\Gamma\to 0$ a small negativity results from this coupling which is given by Eq.~(\ref{mr-38}).

For the VSCPT scheme of laser cooling the dynamical behavior of entanglement is markedly different. In this scheme a free atom is irradiated by counter-propagating orthogonally polarized laser fields. As long as real spontaneous emission events are neglected, the motion of the atom is confined to the three-dimensional manifold $\mathcal{F}(p)$ which is spanned by the product states
\begin{equation}
 \left\{ |3,p\rangle,\, |1,p-k\rangle,\, |2,p+k \rangle \right\},
 \label{dr-4}
\end{equation}
where $p$ denotes the atomic momentum, and $k$ the photon momentum. This fact is a consequence of the conservation of the total momentum of the atom-laser field system, which does not hold for an atom trapped in an external potential. As a result of the finite widths of the dressed atomic states, the atomic wave function is rapidly pumped into the nearly dark state
\begin{equation} \label{chi-p}
 |\chi(p)\rangle = \frac{1}{\sqrt{2}}\left[ \, |1,p-k \rangle
 + |2,p+k \rangle \, \right]
\end{equation}
during the evolution within a particular manifold $\mathcal{F}(p)$. This is a maximally entangled state of the internal and external degrees of freedom of the atom with a negativity of $\frac{1}{2}$.
The effect of spontaneous emissions is that the atomic state vector leaves the manifold $\mathcal{F}(p)$ to enter a new manifold $\mathcal{F}(p')$ in which it is then again driven rapidly to a state of the form (\ref{chi-p}) with the new momentum $p'$.
Within this scheme the cooling proceeds through the consecutive reduction of the atomic momentum to values smaller than that corresponding to the photon momentum, because the atoms are approaching a state which is completely decoupled in the limit $p \to 0$. In fact, in this limit the state
\begin{equation} \label{dark-state}
 |\chi(0)\rangle = \frac{1}{\sqrt{2}}\left[ \, |1,-k \rangle
 + |2,+k \rangle \, \right]
\end{equation}
is reached which is a true dark state that cannot be re-excited by the laser fields. Thus we see that in the long time limit of VSCPT cooling the atom is nearly always in a state of maximal entanglement between the internal and the external degrees of freedom. These states also exist after an instantaneous turnoff of the driving fields. The resulting atomic wavepacket spreads however in space at a rate proportional to the recoil momentum.

\acknowledgments
Financial support by the Deutsche Forschungsgemeinschaft is gratefully acknowledged (Grant No. HE-2525/8).

\end{document}